%% file: ms.tex
\documentclass[fleqn,usenatbib]{mnras}

\usepackage{newtxtext,newtxmath}

\usepackage[T1]{fontenc}

\DeclareRobustCommand{\VAN}[3]{#2}
\let\VANthebibliography\thebibliography
\def\thebibliography{\DeclareRobustCommand{\VAN}[3]{##3}\VANthebibliography}

\usepackage{graphicx}	
\usepackage{amsmath}	
\usepackage{bm}
\usepackage{multirow}

\newcommand{\Lya}{\mathrm{Ly\alpha} }
\newcommand{\HI}{\mathrm{H\textsc{i}}}

\newcommand{\HeII}{\mathrm{He\textsc{ii}}}
\newcommand{\nuL}{\mathrm{\nu_{\rm L}}}

\newcommand{\nHImax}{n_{\rm \HI, max}}
\newcommand{\Msun}{{\rm M}_\odot}

\usepackage[normalem]{ulem}

\title[Emulating extended Ly$\alpha$ haloes]{Emulating extended Lyman-alpha haloes around star-forming galaxies}

\author[Pengfei Li \& Zheng Zheng]{
Pengfei Li\thanks{E-mail: Pengfei.Li@utah.edu}
and Zheng Zheng\thanks{E-mail:
zhengzheng@astro.utah.edu}
\\
Department of Physics and Astronomy, University of Utah, 115 S 1400 E, Salt Lake City, UT 84112, USA
}

\date{Accepted XXX. Received YYY; in original form ZZZ}

\pubyear{2015}

\begin{document}
\label{firstpage}
\pagerange{\pageref{firstpage}--\pageref{lastpage}}
\maketitle

\begin{abstract}
Extended $\Lya$ emission is commonly observed around star-forming galaxies, opening a window for probing the neutral hydrogen gas in the circumgalactic medium (CGM). 
In this paper, we develop a prescription of spherically symmetric CGM gas properties and build emulators to model circularly-averaged surface brightness (SB) profiles of the extended $\Lya$ emission. 
With CGM gas properties parametrized by the density, velocity, and temperature profiles, a self-shielding calculation is carried out to obtain the neutral gas distribution with ionizing photons from the ultraviolet (UV) background and star formation in the galaxy. 
Our calculation reveals three types of systems with distinct neutral gas distribution: non-shielded systems with the CGM being highly ionized across all radii, shielded systems with a neutral gas shell shielding the UV background, and transitional systems in between. 
$\Lya$ SB profiles are obtained through $\Lya$ radiative transfer (RT) simulations, performed for the CGM models with three kinds of $\Lya$ sources: the star formation from central and satellite galaxies, and the recombination in the CGM. 
We build emulators to efficiently predict $\Lya$ SB profiles for given model parameters and $\Lya$ sources, based on Gaussian process regression. 
After being trained with only 180 RT simulations for each $\Lya$ source, the emulators reach an overall accuracy at the level of $\sim 20$ per cent. 
By applying the emulators to fit mock $\Lya$ SB profiles constructed from our model, we find a reasonable recovery of model parameters, indicating the potential of extracting physical information of the CGM and galaxies from the observed extended $\Lya$ emission. 
\end{abstract}

\begin{keywords}
scattering -- galaxies: haloes -- galaxies: high-redshift -- galaxies: intergalactic medium.
\end{keywords}




\input{sec_introduction}

\input{sec_model}
\input{sec_emulator}
\input{sec_application}
\input{sec_summary}

\section*{Acknowledgements}


We thank the anonymous referee for constructive comments. 
This work is supported by NSF grant AST-2007499. 
The support and resources from the Center for High Performance Computing at the University of Utah are gratefully acknowledged. 
P.L. acknowledges support from the Munich Institute for Astro-, Particle and BioPhysics (MIAPbP), which is funded by the Deutsche Forschungsgemeinschaft (DFG, German Research Foundation) under Germany's Excellence Strategy – EXC-2094 – 390783311. 

\section*{Data Availability}
The data underlying this article will be shared on reasonable request to the corresponding author.



\bibliographystyle{mnras}
\bibliography{sec_reference} 



\input{sec_appendix}


\bsp	
\label{lastpage}
\end{document}

%% file: sec_introduction.tex
\section{Introduction} \label{sec:introduction}

The extended $\Lya$ emission is predicted to exist around star-forming galaxies \citep[e.g.][]{Zheng+2011}. 
The discovery of such $\Lya$ haloes (LAHs; e.g. \citealt{Steidel_2011}) has led to numerous follow-up observational and theoretical investigations. 
The resonant interactions between $\Lya$ photons and neutral hydrogen atoms ($\HI$) connect the extended $\Lya$ emission with the underlying gas, providing a valuable window for studying the circumgalactic medium (CGM) of star-forming galaxies. 
In this paper, we aim to build a model for the surface brightness (SB) profiles of the extended $\Lya$ emission and demonstrate its power for constraining physical model parameters by fitting mock profiles.

Because of the low SB of the extended $\Lya$ emission, LAHs were first revealed and detected by stacking narrow-band images of star-forming galaxies 
\citep{Hayashino_2004, Steidel_2011, Matsuda_2012, Feldmeier_2013, Jiang_2013, Xue_2017, Wu_2020, Kakuma_2021, Kikuchihara_2022, Kikuta_2023, Zhang_2024}, over a large redshift range \citep[e.g.][]{Momose_2014, Momose_2016}. They are also observed around individual star-forming galaxies in the local universe, based on effective narrow-band observations from space \citep{Lallement_2011, Hayes_2014, Guaita_2015, Bridge_2018, Runnholm_2023}. 
The extended $\Lya$ emission can be found in spectroscopic observations as well, such as in the long exposure, long-slit spectra \citep{Rauch_2008}. With the commissioning of integral field units (IFU) on large telescopes, LAHs can now be measured and investigated around individual star-forming galaxies \citep{Wisotzki_2016, Wisotzki_2018, Leclercq_2017, Erb_2018, Gallego_2018, Leclercq_2020, Matthee_2020, Kusakabe_2022, Claeyssens_2022, LujanNiemeyer_2022a, LujanNiemeyer_2022b, Guo_2024a, Guo_2024b}.

Interpreting LAHs requires not only understanding the intrinsic production mechanisms of $\Lya$ emission, but also accounting for the interactions between the $\Lya$ emission and the neutral hydrogen gas around galaxies, i.e. the radiative transfer (RT) effects. 

The $\Lya$ production mechanisms mainly include star formation, fluorescence, and cooling 
radiation. The ionizing photons from massive stars produced during star formation can ionize the neutral hydrogen atoms in the interstellar medium, then the subsequent recombination produces $\Lya$ emission \citep{Partridge&Peebles_1967}. 
The ionizing photons from star formation and ultraviolet (UV) background can propagate to the CGM and ionize the neutral hydrogen atoms, followed by the production of $\Lya$ emission through recombination, which is typically called fluorescent $\Lya$ emission \citep[e.g.][]{Haiman_2001, Kollmeier_2010}. 
Finally, $\Lya$ photons can be produced as part of the cooling radiation from the gas accreted from the intergalactic medium (IGM) and collisionally excited in the CGM \citep{Dijkstra_2009, FaucherGiguere_2010, Goerdt_2010, Cantalupo_2014, Rosdahl_2012,Ledos_2023}. 

After $\Lya$ photons are produced, they experience a large number of scatterings with neutral hydrogen atoms in the CGM, which causes diffusion in both space and frequency. Such an RT process depends on the density, velocity and temperature of the neutral hydrogen gas, which requires theoretical modelling for properly interpreting LAHs.

Various theoretical frameworks have been explored to understand LAHs, ranging from analytically solving RT equations to numerical RT simulations. 
Analytical solutions of $\Lya$ SB profiles and spectra are obtained for a 
power-law gas density profile in the optically thick regime \citep{Lao_2020}. 
A more flexible framework is to perform Monte Carlo RT simulations with an analytical setup of gas properties, such as multi-phase gas with clumpy and diffuse 
components or outflowing gas \citep[e.g.][]{Dijkstra_2012, Gronke_2016, Song_2020, Chang_2023}. 
Such models have been successfully applied to fit $\Lya$ spectra and SB profiles of observed LAHs \citep[e.g.][]{Song_2020, Erb_2023, Garel_2024}. Another framework is to carry out Monte Carlo RT calculations for galaxies in hydrodynamic galaxy formation simulations \citep{Laursen_2007, Zheng_2010, Verhamme_2012}, which can generally produce $\Lya$ SB profiles similar to observations \citep{Lake_2015, Smith_2019, Byrohl_2021, Mitchell_2021}. 

In this paper, we construct a model of extended LAHs, aiming to interpret observed $\Lya$ SB profiles. It captures the key physical components, including multiple sources of $\Lya$ production and RT in the CGM. Given that RT is computationally expensive, we develop emulators to improve the efficiency of modelling extended LAHs. 

We organize the paper as follows. In Section \ref{sec:GasModel}, we present the model for the neutral hydrogen gas, the $\Lya$ 
sources, and the calculation of $\Lya$ SB profiles from RT simulations.
In Section \ref{sec:Emulator}, the design and construction of emulators for $\Lya$ SB profiles are introduced and the performance of emulators is examined and assessed. We apply the emulators to model mock $\Lya$ SB profiles in Section \ref{sec:application} to demonstrate the potential of extracting CGM and galaxy properties from observations of LAHs. 
In Section \ref{sec:summary}, we summarize the work and discuss possible improvements.

Throughout this paper, whenever necessary, we adopt a spatially-flat $\Lambda \rm CDM$ cosmological model consistent with the constraints from \textit{WMAP5} \citep{Komatsu_2009}, which has $\Omega_{\rm m}=0.28$, $\Omega_{\rm b}=0.046$, $\sigma_8=0.82$, $n_{\rm s}=0.96$, and $H_0=100h\, {\rm km\,s^{-1}Mpc^{-1}}$ with $h=0.7$. The logarithm $\log$ in this paper is base 10.

%% file: sec_model.tex
\section{Model Description} \label{sec:GasModel}

In this section, we describe our model of $\Lya$ SB profiles around star-forming galaxies. There are three major components: neutral hydrogen ($\HI$) gas 
properties, intrinsic $\Lya$ sources, and $\Lya$ RT simulations. The CGM in our model is assumed to be spherically symmetric, described by density, velocity, temperature, and ionizing fraction profiles. The ionizing fraction profile is obtained through a self-shielding calculation. The intrinsic $\Lya$ sources include star formation from central galaxies, CGM recombination, and 
star formation from satellite galaxies. With gas properties and $\Lya$ sources, we perform RT simulations to produce $\Lya$ SB profiles.

\subsection{Gas properties} \label{sec:GasProperties}

We model gas density closely following \cite{Padmanabhan+2016}. The total gas mass $M_{\rm g}$ within the viral radius is determined by the host halo mass $M_{\rm h}$ \citep{Barnes_2010}, 
\begin{equation}
    M_{\rm g} = f_{\rm b} M_{\rm h} \exp{\left[- \left( \frac{v_{\rm c, 0}}{v_{\rm c}(M_{\rm h})} \right)^3 \right]},
    \label{eq:HydrogenMass}
\end{equation}
where $f_{\rm b}=\Omega_{\rm b}/ \Omega_{\rm m}$ is the global baryon fraction, $v_{\rm c}(M_{\rm h})$ 
is the circular velocity at the viral radius $R_{\rm vir}$, and $v_{\rm c, 0} = 50 \rm km\, s^{-1}$. 
Halo mass is defined to have a mean matter density (within $R_{\rm vir}$) of $18\pi^2$ times the critical density at a given redshift. 

Following \cite{Maller_2004}, the gas density profile is modelled with a modified Navarro-Frenk-White (NFW) profile \citep{NFW_1996} to account for the baryonic effects,  
\begin{equation}
    \rho (r) = \frac{\rho_{\rm 0}}{(c_{\rm g} r/R_{\rm vir} + 0.75)(c_{\rm g}r/R_{\rm vir} + 1)^2}.
    \label{eq:ModifiedNFW}
\end{equation}
The gas concentration parameter $c_{\rm g}$ is modelled as \citep{Maccio_2007} 
\begin{equation}
    c_{\rm g} = c_{\rm 0} \left( \frac{M_{\rm h}}{10^{11} \Msun} \right)^{-0.109} \frac{4}{1 + z}, 
    \label{eq:c0}
\end{equation}
where $c_{\rm 0}$ is a free parameter and $z$ is the redshift.
With the gas concentration and the total gas mass [equation~(\ref{eq:HydrogenMass})], $\rho_0$ is determined through 
\begin{equation}
    M_{\rm g} =  \int_0^{R_{\rm vir}} \rho (r) 4 \pi r^2 dr  .
    \label{eq:FixRho0}
\end{equation}
We further add the mean gas density $\bar{\rho}_{\rm b}$ of the universe
to equation~(\ref{eq:ModifiedNFW}) to describe the IGM at large radii, and the total gas density profile is 
\begin{equation}
    \rho_{\rm g} (r) = \rho (r) + \bar{\rho}_{\rm b}.
    \label{eq:rhog}
\end{equation}
The dashed curve in the top panel of Fig.~\ref{fig:ssc} shows one example of the total hydrogen density profile, which is the total gas density profile times the hydrogen mass abundance 0.76. 

The radial velocity profile of the gas is modelled following \cite{Sadoun_2017} as 
\begin{equation}
    v (r) = \frac{v_{\rm B}(r)}{1 + e^{-20(r/R_{\rm vir}- 1)}}. 
    \label{eq:velocity}
\end{equation}
The quantity $v_{\rm B}(r)$ is the average radial velocity around a dark matter halo of mass $M_{\rm h}$ at redshift $z$, calculated based on the extended Press–Schechter formalism in \cite{Barkana2004}. 
Outside of the viral radius, this form of $v(r)$ essentially follows $v_{\rm B}(r)$, which appears as infall close to the halo and approaches to Hubble flow at large radii. 
Within the halo viral radius, we assume that the gas only has random motion, and therefore modify the velocity profile $v_{\rm B}(r)$ as in equation~(\ref{eq:velocity}) such that the radial velocity $v(r)$ approaches to zero. 
In reality, the velocity distribution can be much more complex with inflow, outflow and turbulence. For simplicity, we use the above velocity profiles to demonstrate our framework. One example of the velocity profile is shown in the bottom panel of Fig.~\ref{fig:ssc}. 

The gas temperature is relevant for two calculations. For the self-shielding calculation (Section \ref{sec:SSC}), we fix it to be $10^4 \rm K$, around the peak of the cooling rate. For the RT calculation (Section \ref{sec:RT}), we fix the temperature to be $10^6 \rm K$ to account for the gas random motion (e.g. turbulence). 

For our later calculations with the model, we set the boundary of the gas distribution to be $10R_{\rm vir}$. This is motivated by several reasons. First, if the virial radius of $10^{11} M_\odot$ haloes at $z\sim 3$ is used as a reference, $\Lya$ haloes around individual galaxies are detected at a radius beyond the halo virial radius \citep[e.g.][]{Leclercq_2017}. For $\Lya$ haloes detected through stacking narrow-band images, they can even reach a few times the virial radius \citep[e.g.][]{Momose_2016,LujanNiemeyer_2022a}. Next, the neutral gas outside the virial radius (IGM) can still affect the $\Lya$ RT, based on theoretical investigations \citep[e.g.][]{Dijkstra_2007,Zheng_2011a} and observational study \citep[e.g.][]{Momose_2021}. In particular, a significant fraction of scatterings can occur in the infall region outside of the virial radius, which influence the spectral and spatial distribution of $\Lya$ photons \cite[e.g.][]{Sadoun_2017}. 

\subsection{Self-shielding calculation} \label{sec:SSC}

$\Lya$ photons interact with neutral hydrogen atoms. Given the gas profile in equation~(\ref{eq:rhog}), we solve the gas ionization structure by performing a self-shielding calculation. We consider two types of ionizing sources, the star formation in central galaxies and the UV ionizing background. 

For the ionizing photons from star formation, the production rate $Q$ is related to the star formation rate (SFR) as 
\begin{equation}
\begin{split}
    \log (Q/ {\rm s^{-1}}) =\ & 53.8 + \log [{\rm SFR}/(\Msun  {\rm yr^{-1}})] \\
                - &\ 0.0029[9 + \log (Z/Z_\odot)]^{2.5}, 
    \label{eq:IonizingPhotons}
\end{split}
\end{equation}
based on the Salpeter initial mass function and a constant SFR \citep{Schaerer_2003, Dijkstra_2007}.
We fix the metallicity $Z$ to be solar metallicity ($Z/Z_\odot=1)$.
Following \cite{Zheng_2010}, we connect SFR to halo mass, 
\begin{equation}
    {\rm SFR}/(\Msun {\rm yr^{-1}}) = 0.68 [M_{\rm h}/(10^{10} h^{-1}\Msun)].
    \label{eq:SFR}
\end{equation}
The frequency-dependent luminosity $L_{\nu}$ of ionizing photons is assumed to 
follow a power law, $L_{\nu} = \beta Q h_{\rm p} (\nu/\nuL)^{-\beta}$, with the normalization 
coming from $Q=\int_{\nuL}^\infty L_\nu/(h_{\rm p}\nu) d\nu$, 
where $h_{\rm p}$ is the Plank constant, and $\nuL$ is the $\HI$ Lyman-limit frequency. 
Following \cite{Sadoun_2017}, we fix $\beta=3$.

For the UV ionizing background, we adopt the model in \cite{Haardt_2012} to obtain the redshift-dependent
$\HI$ photoionization rate $\Gamma_{\rm UV}$. We assume a constant intensity 
$I_{\nu, 0}$ between $\nuL$ and $4\nuL$ (the $\HeII$ Lyman-limit), which is derived from 
$\Gamma_{\rm UV} = 4\pi \int_\nuL^{4\nuL} I_{\nu,0} / (h_{\rm p} \nu) a_\nu d\nu$, 
with $a_\nu$ being the photoionization cross section.

With the two ionizing sources, we perform a self-shielding calculation to determine the neutral 
hydrogen fraction $x_{\rm \HI} (r) \equiv n_{\rm \HI}(r)/n_{\rm H}(r)$ as a function of radius $r$, where 
$n_\HI$ and $n_{\rm H}$ are the number densities of neutral and total hydrogen, respectively. 
This is achieved by solving the photoionization equilibrium equation 
\begin{equation}
 x_{\rm \HI}(r) \Gamma_{\rm tot, ss}(r) = \alpha_{\rm B}(T) [1 - x_{\rm \HI}(r)]n_{\rm e}, 
    \label{eq:PhotoIonEq}
\end{equation}
where $\alpha_{\rm B}$ is the case-B recombination coefficient at temperature $T$. The electron number density is calculated as $n_{\rm e} = 0.82 (1-x_{\rm \HI}) \rho_{\rm g}/m_{\rm H}$, accounting for the contribution from singly ionized helium, with $m_{\rm H}$ being the mass of a hydrogen atom. The total photoionization rate $\Gamma_{\rm tot, ss}$ comes from the attenuated ionizing radiation from the central star formation and the UV background, $\Gamma_{\rm tot, ss} = \Gamma_{\rm cen, ss} + \Gamma_{\rm UV, ss}$, with 
\begin{equation}
\begin{split}
    \Gamma_{\rm cen, ss} (r) = & \int_{\nuL}^{\infty} \frac{f_{\rm esc} L_\nu \exp [-\tau_{\rm \nu,cen} (r) ] } {4 \pi r^2 h_{\rm p} \nu} a_{\nu} d\nu 
    \label{eq:IonSourceCen}
\end{split}
\end{equation}
and 
\begin{equation}
\begin{split}
    \Gamma_{\rm UV, ss} (r) = & \int_{4 \pi} d\Omega \int_{\nuL}^{\infty} \frac{I_{\nu, 0} \exp [-\tau_{\rm \nu, UV} (r,\bm{\hat{n}})]}{h_{\rm p} \nu} a_\nu d\nu.
    \label{eq:IonSourceUV}
\end{split}
\end{equation}
In equation~(\ref{eq:IonSourceCen}), $f_{\rm esc}$ is the escaping fraction of central ionizing photons, and $\tau_{\rm \nu, cen}(r)$ is the photoionization optical depth from the centre to radius $r$. 
In equation~(\ref{eq:IonSourceUV}), $\tau_{\rm \nu, UV} (r,\bm{\hat{n}})$ is the photoionization optical depth from $10 R_{\rm vir}$ to radius $r$ along direction $\bm{\hat{n}}$.  

\begin{figure}
	\includegraphics[width=\columnwidth]{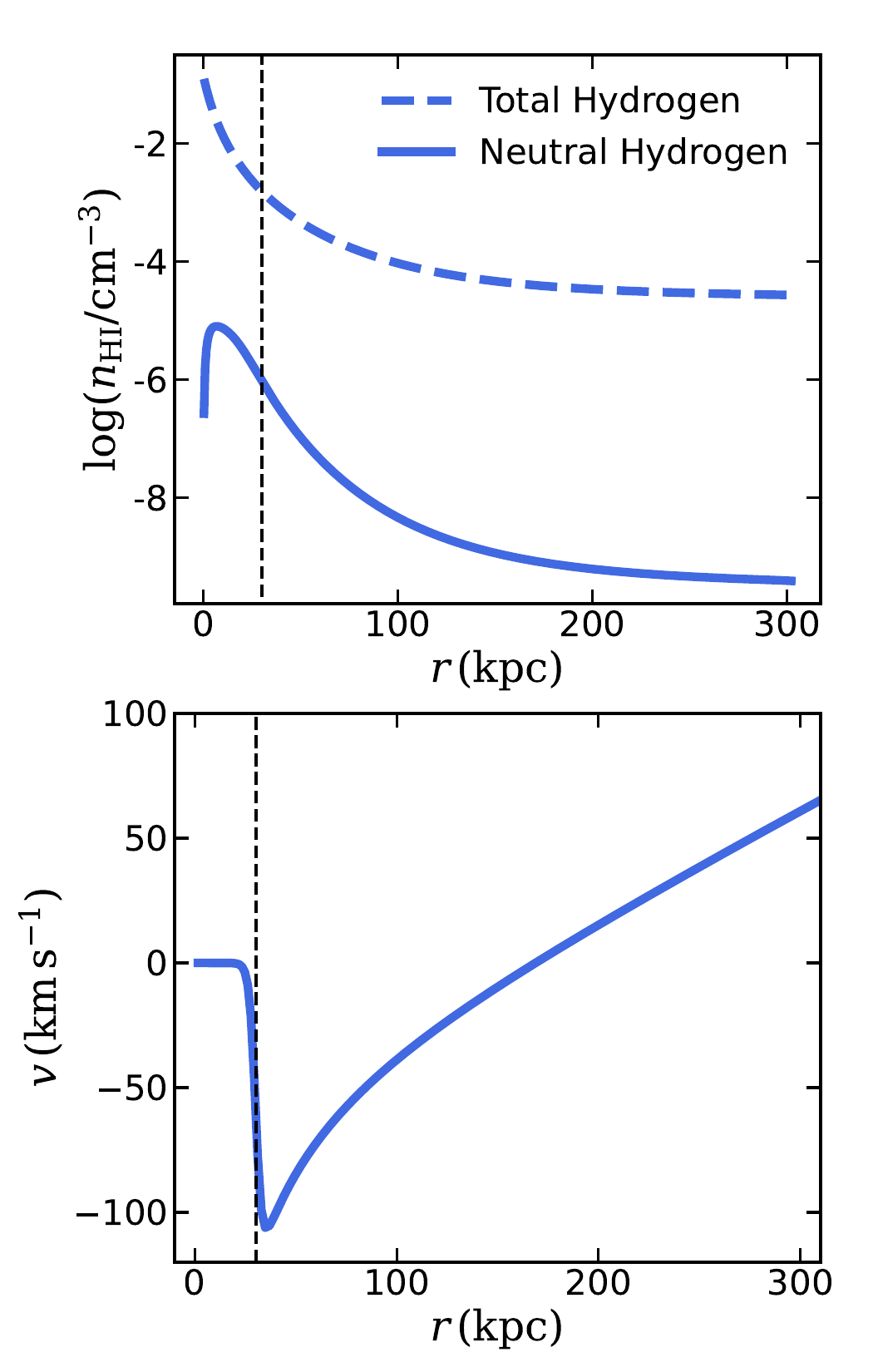}
    \caption{Profiles of gas properties. Top: The dashed curve is the total hydrogen gas density profile. The solid curve is the neutral hydrogen density profile obtained through the self-shielding calculation by accounting for ionizing photons from the central source and the UV background. 
    Bottom: The gas velocity profile composed of the dispersion-dominated part within $R_{\rm vir}$, the infall region outside of $R_{\rm vir}$, and the Hubble flow at larger radii. In both panels, the vertical line shows the $R_{\rm vir}$. The model parameters for this system are $\log (M_{\rm h}/\Msun)=11$, $z=4$, 
    $\log c_{\rm 0}=0.6$, and $\log f_{\rm esc}=-1$. 
    The radius $r$ is in units of physical kpc.}
    \label{fig:ssc}
\end{figure}

The self-shielding calculation is performed in an iterative way \citep{Zheng+2002b}. We initialize $x_\HI (r)$ under the optically thin condition (i.e. $\tau_{\rm \nu, cen}=0$ and $\tau_{\rm \nu, UV}=0$). We then recompute the optical depths by integrating the $\HI$ density profile, and update $x_{\rm \HI}$ using equation~(\ref{eq:PhotoIonEq}).
This process is repeated until the maximum fractional difference of $x_\HI$ between two 
consecutive iterations is smaller than $10^{-5}$. The solid curve in the top panel of Fig.~\ref{fig:ssc} shows one example of the neutral hydrogen density profile after the self-shielding calculation.

\subsection{Intrinsic $\Lya$ sources} \label{sec:LyaSource}

There are three intrinsic $\Lya$ sources in our model, star formation in the central galaxy, recombination in the CGM, and star formation in satellite galaxies, denoted as central, CGM, and satellite sources, respectively. 
In principle, besides the recombination, the collisional de-excitation of neutral hydrogen atoms following the collisional excitation can lead to another extended $\Lya$ source in the CGM. 
Because its modelling procedure is similar to the recombination in the CGM, for our purpose of proof of concept, we choose not to consider it in this paper. 

For the central source, the intrinsic $\Lya$ luminosity is related to $\rm SFR$ following \cite{Zheng_2010}, 
\begin{equation}
    L_{\rm cen}/({\rm erg\, s^{-1}}) = (1 - f_{\rm esc}) 10^{42} [ {\rm SFR}/ ( \Msun\, {\rm yr^{-1}} )] ,
    \label{eq:LyaCentral}
\end{equation}
where the $1-f_{\rm esc}$ factor accounts for the escaping of ionizing photons from the star-forming region. 
For the CGM source, the $\Lya$ emissivity at each radius is estimated to follow two-thirds of the recombination rate \citep{Dijkstra_2014}, $\epsilon(r)=(2/3)\alpha_{\rm B}n_{\rm e} n_{\rm H} (1-x_{\HI})$. 
For the satellite source, we describe it as a galaxy located at radius $R_{\rm sat}$ with $\Lya$ luminosity $L_{\rm sat}$. 

\begin{figure*}
	\includegraphics[width=\textwidth]{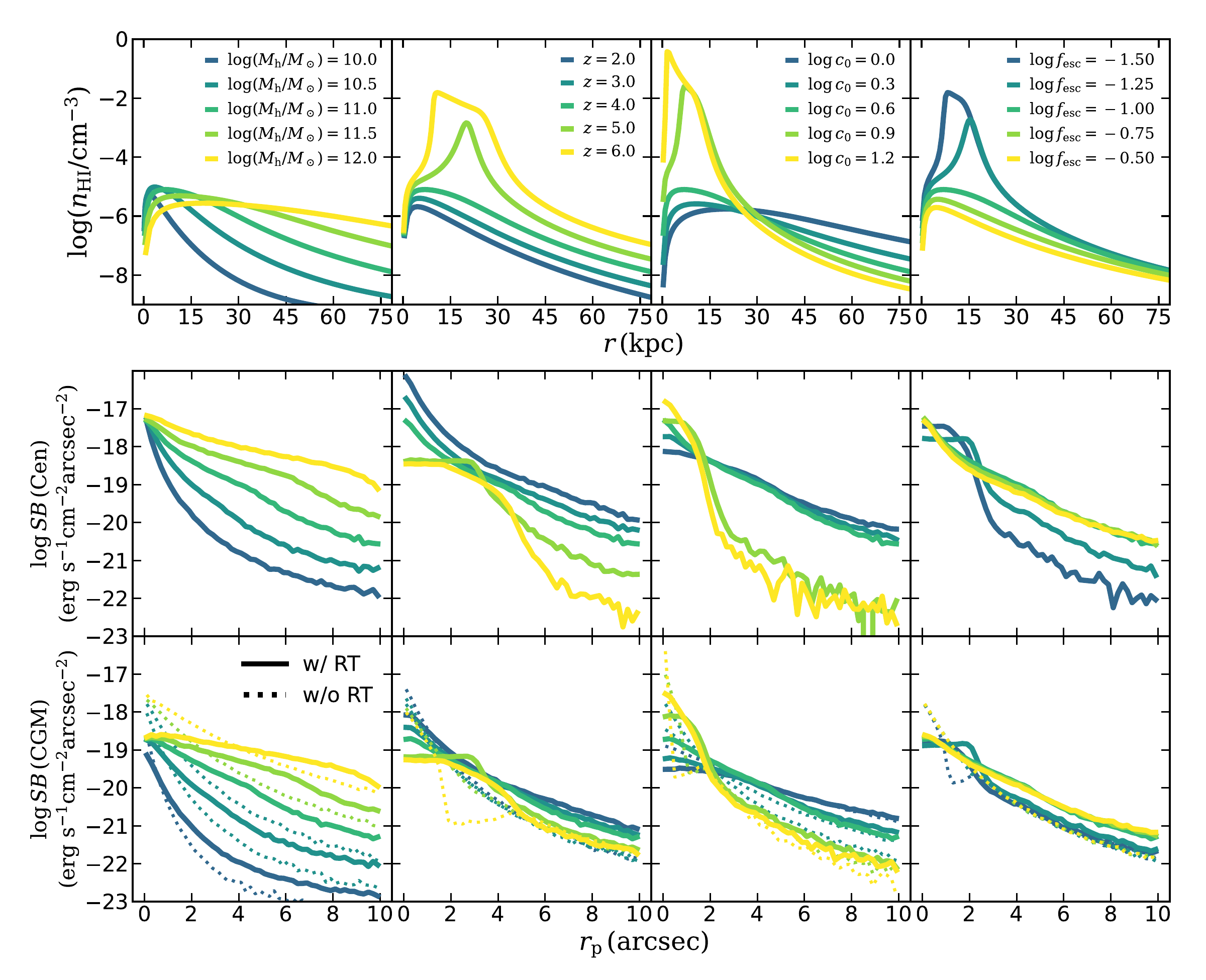}
    \caption{Neutral hydrogen density profiles and $\Lya$ SB profiles for different gas configurations and $\Lya$ sources. The first row shows neutral hydrogen density profiles. The second and the third rows show $\Lya$ SB profiles for central and CGM sources, respectively. In the bottom row, the intrinsic $\Lya$ SB profiles without RT (dotted lines) are shown, as a comparison to those with RT (solid lines). From a base model with $\log (M_{\rm h}/\Msun)=11$, $z=4$, $\log c_{\rm 0}=0.6$, and $\log f_{\rm esc}=-1$, each of the four columns corresponds to varying one model parameter while keeping the rest parameters fixed. The radius $r$ in the top row is in units of physical kpc, while the projected radius $r_{\rm p}$ in the bottom two rows is put in units of arcsecond. 
    }
    \label{fig:trend_cc}
\end{figure*}

\subsection{$\Lya$ RT and model $\Lya$ SB profiles} \label{sec:RT}

To calculate $\Lya$ SB profiles for given gas models and intrinsic $\Lya$ sources, we run RT simulations adopting the method in \cite{Zheng+2002a}. With each $\Lya$ source (central, CGM, or satellite), we run the RT simulation with $10^5$ photons for each training or testing model of the emulators. For each source, the photons are equally weighted to reproduce the total $\Lya$ luminosity of the source. 

First, we initialize $\Lya$ photons at the line centre in the lab frame\footnote{
In principle, it is more physical to initialize the frequency at the line centre in the rest frame of the atom, resulting in a thermally broadened (Gaussian) distribution in the lab frame. In practice, this initialization and our initialization are almost equivalent in terms of the subsequent frequency distribution of $\Lya$ photons in the lab frame. The reason is that the bulk gas velocity in our model is typically smaller than the thermal velocity (with one-dimensional dispersion of $\sim 130\, {\rm km\, s^{-1}}$ for the assumed $T=10^6 {\rm K}$). The first a few scatterings encountered by the $\Lya$ photons would bring them toward the line centre in the rest frame of atoms, giving rise to a thermally broadened frequency distribution in the lab frame. 
}
for different sources. 
The $\Lya$ photons of central and satellite sources are launched at the centre of the system and at the position of the satellite (radius $R_{\rm sat}$), respectively. 
While for the CGM source,  to determine the launching position of each photon, a random radial direction is chosen and then the radius  is sampled according to the probability distribution of the volume-weighted $\Lya$ emissivity, $P(r)\propto \epsilon(r)r^2$. 
Then we trace each photon's propagation until it escapes the system at $10 R_{\rm vir}$, after which we record its frequency, propagating direction, and last scattering position. 
Every scattering is considered in our RT calculation, with no acceleration scheme applied. 

For central and CGM sources, given the spherical symmetry, we can split photons into different projected radius $r_{\rm p}$ bins. For photons with total luminosity $L_{\rm A}$ within a radial bin of annulus area $A$ centred at $r_{\rm p}$, the surface brightness is calculated as \citep{Lake_2015} 
\begin{equation}
    {\rm SB}(r_{\rm p}) =  \frac{L_{\rm A}}{4\pi A (1+z)^4}.
    \label{eq:SBP_Cen}
\end{equation}

For satellite sources, we choose the radial direction passing the satellite as the $z$-axis, with the origin at the centre of the system. 
With a satellite at a given position, we collect photons within a narrow bin of an observing direction $\mu \pm \Delta \mu/2$, where $\mu$ is the cosine of the angle between the direction towards the observer and the $z$-axis. Then we calculate the circularly averaged surface brightness profiles following equation~(\ref{eq:SBP_Cen}). 
Each satellite $\Lya$ SB profile corresponds to only one satellite source in a system. For modelling simplicity, satellite sources do not have their own gaseous halos and make no contribution to ionizing photons. 

Throughout this paper, we take the bin size of the projected radius to be $\Delta r_{\rm p}=0.2\ \rm arcsec$ (motivated by observations; e.g. \citealt{Leclercq_2017}), and the angular bin size of satellite sources to be $\Delta \mu=0.1$. 

Fig.~\ref{fig:trend_cc} shows the response of $\Lya$ SB profiles to different model parameters for central and CGM sources, along with neutral hydrogen density profiles. 
For columns from left to right, the parameters shown are halo mass $\log M_{\rm h}$, redshift $z$, gas concentration parameter $\log c_0$, and the escaping fraction of ionizing photons $\log f_{\rm esc}$. We start from a model with $\log (M_{\rm h}/\Msun)=11$, $z=4$, $\log c_{\rm 0}=0.6$, and $\log f_{\rm esc}=-1$. 
Then we vary one model parameter at a time for each column. 
As a comparison between $\Lya$ SB profiles with and without RT, we plot the intrinsic $\Lya$ SB profiles of the CGM source in the bottom row with dotted lines. RT results in spatial diffusion of $\Lya$ photons, which explains the subsequent, more extended $\Lya$ SB profiles, with sharp features towards the centre smoothened. Such an RT-induced smoothing effect facilitates model interpolations and emulator construction. 
For the last three columns, a clear bimodality can be seen for both $\Lya$ SB profiles and neutral hydrogen density profiles. 
One type has low neutral hydrogen densities across all radii and extended $\Lya$ SB profiles. 
These systems have either a stronger UV ionizing background at lower $z$, a less concentrated gas density profile (smaller $\log c_0$), or a stronger photoionization effect from the central source (larger $\log f_{\rm esc}$). 
The other type has strongly peaked neutral density profiles near the centre and nearly flat $\Lya$ SB profiles around the centre followed by a sharp drop, for reasons opposite to the previous case. 
We call these two types of systems as \textit{non-shielded} and \textit{shielded} systems for later references. 
The $\Lya$ photons in shielded systems experience significantly more scattering events in the inner region compared to those in non-shielded systems, leading to a stronger frequency diffusion away from the line centre and an easier escape from smaller radii. 
Such RT processes lead to much steeper $\Lya$ SB profiles in shielded systems than those in non-shielded systems. 
The transitions between these two types of systems are discussed in Appendix \ref{sec:transition}. 
The system type depends on the combination of model parameters. For the specific parameter combinations in the first column, only non-shielded systems are present. 

\begin{figure}
    \includegraphics[width=\columnwidth]{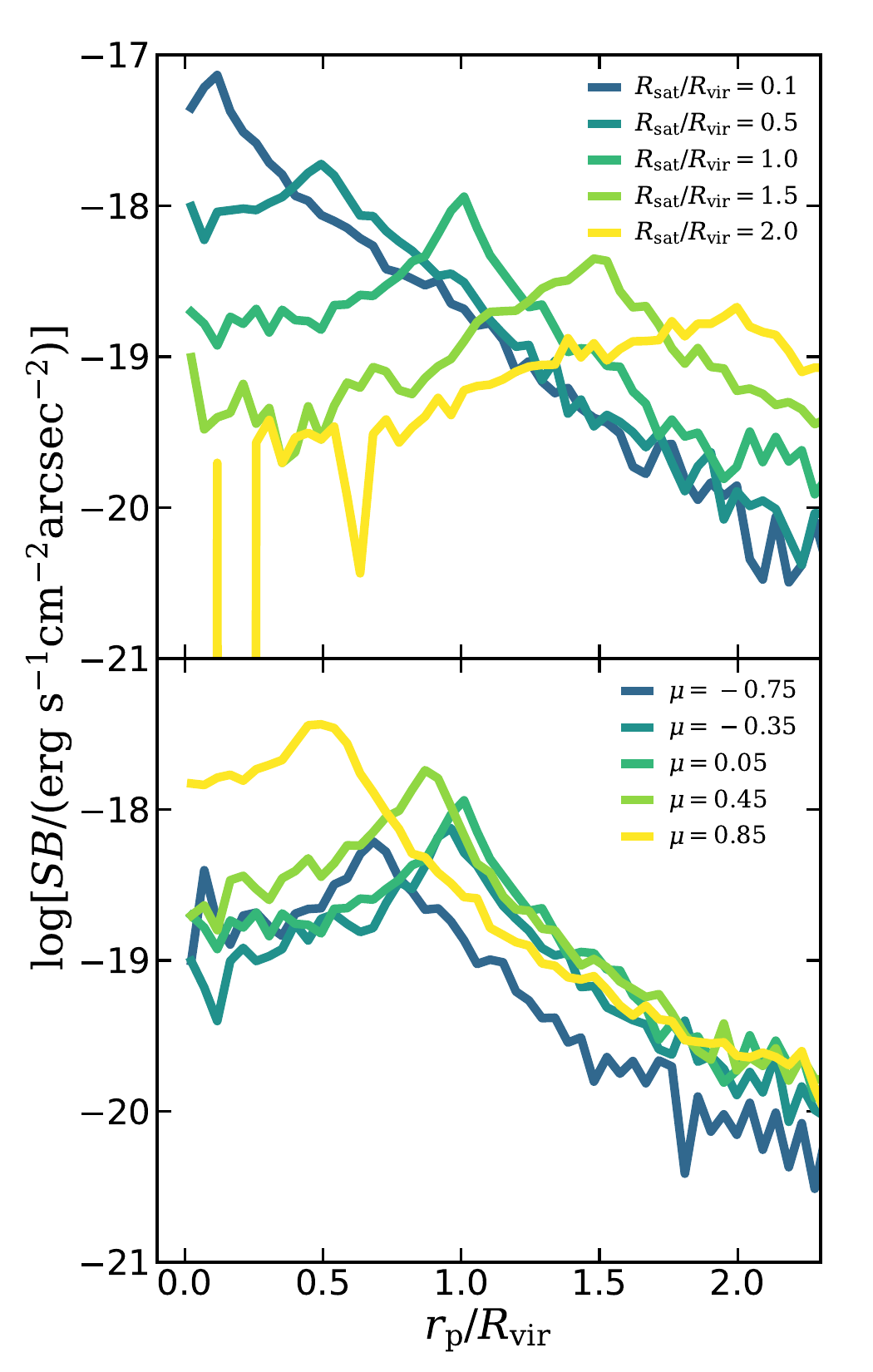}
    \caption{Circularly averaged $\Lya$ SB profiles for satellite sources. From a model with $\log (M_{\rm h}/\Msun)=11$, $z=4$, $\log c_{\rm 0}=0.6$, and $\log f_{\rm esc}=-1.0$, the top panel shows satellite $\Lya$ SB profiles for different satellite positions $R_{\rm sat}/R_{\rm vir}$ (with $\mu=0.05$), and the bottom panel shows satellite $\Lya$ SB profiles for different observing directions $\mu$ (with $R_{\rm sat}/R_{\rm vir}=1$). The peaks in the $\Lya$ SB profiles indicate the projected positions of satellites.}
    \label{fig:trend_sate}
\end{figure}

Fig.~\ref{fig:trend_sate} shows the circularly averaged $\Lya$ SB profiles for satellite sources, which are put on the $z$-axis. With fixed gas distribution and satellite luminosity, we show how $\Lya$ SB profiles vary with the two parameters related 
to satellite galaxies: the distance of the satellite to the system centre, $R_{\rm sat}$, and the observing direction with respect to $z$-axis, $\mu$. The top panel shows the dependence of $\Lya$ SB profiles
on satellite position $R_{\rm sat}$, with fixed $\mu=0.05$. The profile peaks around the projected position of a satellite. 
When the satellite is put at a larger radius, the overall amplitude decreases, with the profile becoming more extended. 
The bottom panel shows the dependence of $\Lya$ profiles on the observing direction $\mu$, with fixed $R_{\rm sat}/R_{\rm vir}=1.0$. 
With respect to the system centre, if the satellite is on the near side towards the observer ($\mu>0$), the amplitude of the $\Lya$ SB profile is higher with a more clear peak, compared to the satellite on the far side ($\mu<0$). 
This is expected because $\Lya$ photons from the far side experience a larger $\HI$ column density and more spatial diffusion before reaching the observer. 

To compare with observed $\Lya$ SB profiles, we need to account for the seeing effect. We process this 
before calculating $\Lya$ SB profiles by redistributing simulated photons following a two-dimensional (2D) Gaussian point spread function (PSF) characterised by the full width at half maximum (FWHM). For a simulated photon with projected radius $r_{\rm p}$, we draw two random numbers, $\Delta x$ and $\Delta y$, following the 2D Gaussian distribution, and 
update $r_{\rm p}$ to $\sqrt{(r_{\rm p} + \Delta x)^2 + \Delta y^2}$. Then we 
calculate $\Lya$ SB profiles following equation~(\ref{eq:SBP_Cen}). The typical FWHM from the MUSE observation is about $0.7\ \mathrm{arcsec}$ \citep[e.g.][]{Leclercq_2017}, which is 
adopted for presenting results in the rest of this paper.

%% file: sec_emulator.tex
\section{Construction of the Emulators} \label{sec:Emulator}

To model the observed $\Lya$ SB profiles, we need to explore the parameter space and run RT simulations for each model, which would be time-consuming. To make the parameter exploration more efficient, we adopt the emulator technique, which builds a fast mapping from model parameters to $\Lya$ SB profiles. We describe the design of the emulators and their performance in this section. 

\subsection{Latin hypercube design} \label{sec:SLHD}
To build an emulator, we first need to run RT simulations to obtain $\Lya$ SB profiles at a number of positions in the parameter space, and convolve them with the adopted PSF. 
The chosen parameters need to have a good coverage of the entire parameter space. 
Additionally, we want the number of these positions to be small to reduce the number of RT simulations. 
These two requirements can be achieved through the Latin hypercube design (LHD). 
The first step of an LHD is to discretize a $d$ dimensional space into $n^d$ cubes uniformly, where $n$ is the number of grids for each dimension. 
Then $n$ points are populated into the discretized space, with each row of any dimension having one point and one point only. 
In this work, $n$ is the number of models for which we run RT simulations and $d$ is the number of model parameters.

To generate an LHD with given $n$ and $d$, we need to draw $n$ points in a $d$-dimensional space. The coordinates of the $i$-th point is $\left(X_1^{(i)}, X_2^{(i)}, ..., X_d^{(i)}\right)$, whose component is denoted as $X_k^{(i)}$ with $k=1$, $2$, ..., and $d$. For a given $k$, the components $X_k^{(i)}$ of the $n$ points are drawn from the permutation of $1$, $2$, ..., and $n$. 
A simple example of an LHD is to distribute these points diagonally in this $d$ dimensional space. 
However, such a design does not cover the parameter space effectively, and an optimization algorithm needs to be applied for better coverage. 

We first randomly initialize a $d$-dimensional LHD with $n$ points. 
For each pair of points, we exchange their $k$-th coordinate components and calculate the entropy of the LHD after the exchange. 
Only the LHD with the largest entropy is kept among these exchanges. 
The above process is repeated from $k=1$ to $k=d$. 
Then the process is iterated until the fractional difference of entropies between two consecutive iterations is smaller than 0.01.
The entropy of an LHD is defined as $\ln |S|$, with the matrix $S$ having elements 
\begin{equation}
    S_{ij} = \exp \left[-\sum_{k=1}^d \left(x_k^{(i)} - x_k^{(j)} \right)^2 \right],
    \label{eq:entSLHD}
\end{equation}
where $x_k^{(i)}=X_k^{(i)}/n$ \citep{Ye_2000}, and $|S|$ is the determinant of the matrix $S$. 
We further require the parameter distribution to be symmetric about the centre of the parameter space during both the initialization and optimization processes to speed up the optimization, which is called the symmetric Latin hypercube design  (SLHD; \citealt{Ye_2000}). 

\begin{figure}
	\includegraphics[width=\columnwidth]{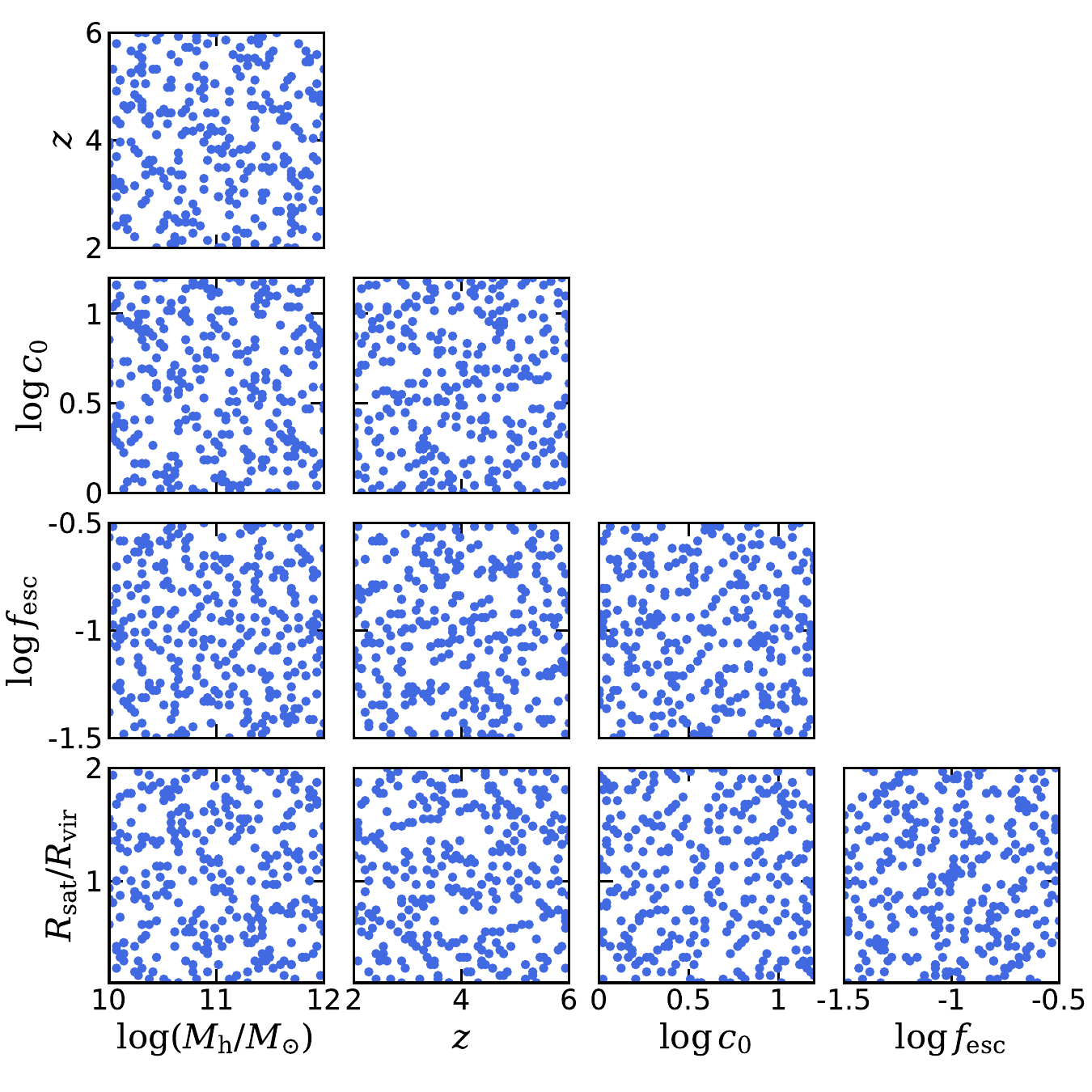}
    \caption{Parameter distribution for the training models, designed through the symmetric Latin hypercube technique with 180 models.}
    \label{fig:SLHD}
\end{figure}

We design three $n \times d = 60 \times 5$ Latin hypercubes (60 models with 5 parameters) and combine them. These $N=180$ models are the training models for developing emulators, and we run RT simulations for each of them. 
The five parameters are $\log M_{\rm h}$, $z$, $\log c_{\rm 0}$, $\log f_{\rm esc}$, and $R_{\rm sat}/R_{\rm vir}$. 
The design result is shown in Fig.~\ref{fig:SLHD}, which displays a uniform coverage of the parameter space for the five parameters. 
For central and CGM sources, we only need the first four parameters. 
For satellite sources, we need all five parameters. 
The average CPU (2.1 GHz) hours for finishing 180 RT simulations of central, CGM, and satellite sources are about 46k, 34k, and 7k, respectively.

\subsection{Gaussian process regression} \label{sec:GPR}
\begin{figure*}
	\includegraphics[width=\textwidth]{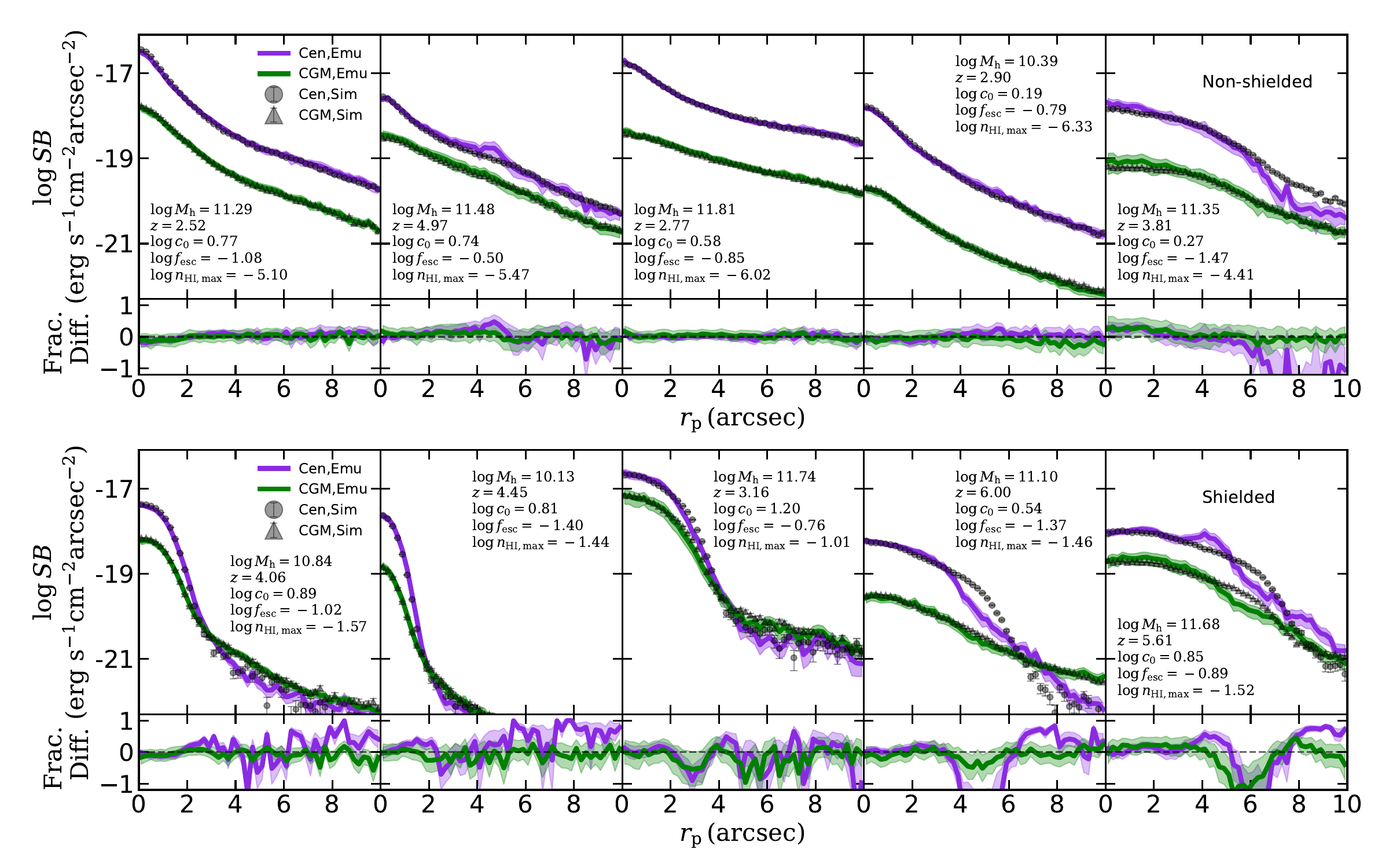}
    \caption{Examples of emulator performance for central and CGM sources in non-shielded systems (top row) and shielded systems (bottom row). 
    In the top panels of each row, the grey points and error bars show $\Lya$ SB profiles and the shot noise, respectively, from RT simulations. 
    The solid lines show emulator predictions (means). 
    The shaded regions show the uncertainties predicted by emulators. 
    The model parameters for each system are listed in the panel ($M_{\rm h}$ in units of $\Msun$; $\nHImax$ in units of $\rm cm^{-3}$). 
    In the bottom panels of each row, the solid lines show the fractional differences of $\Lya$ SB profiles from emulator predictions and from RT simulations, with respect to emulator predictions. 
    The shaded region shows uncertainties predicted by emulators in terms of the fractional difference.}
    \label{fig:slhdtest_cc}
\end{figure*}

\begin{figure*}
	\includegraphics[width=\textwidth]{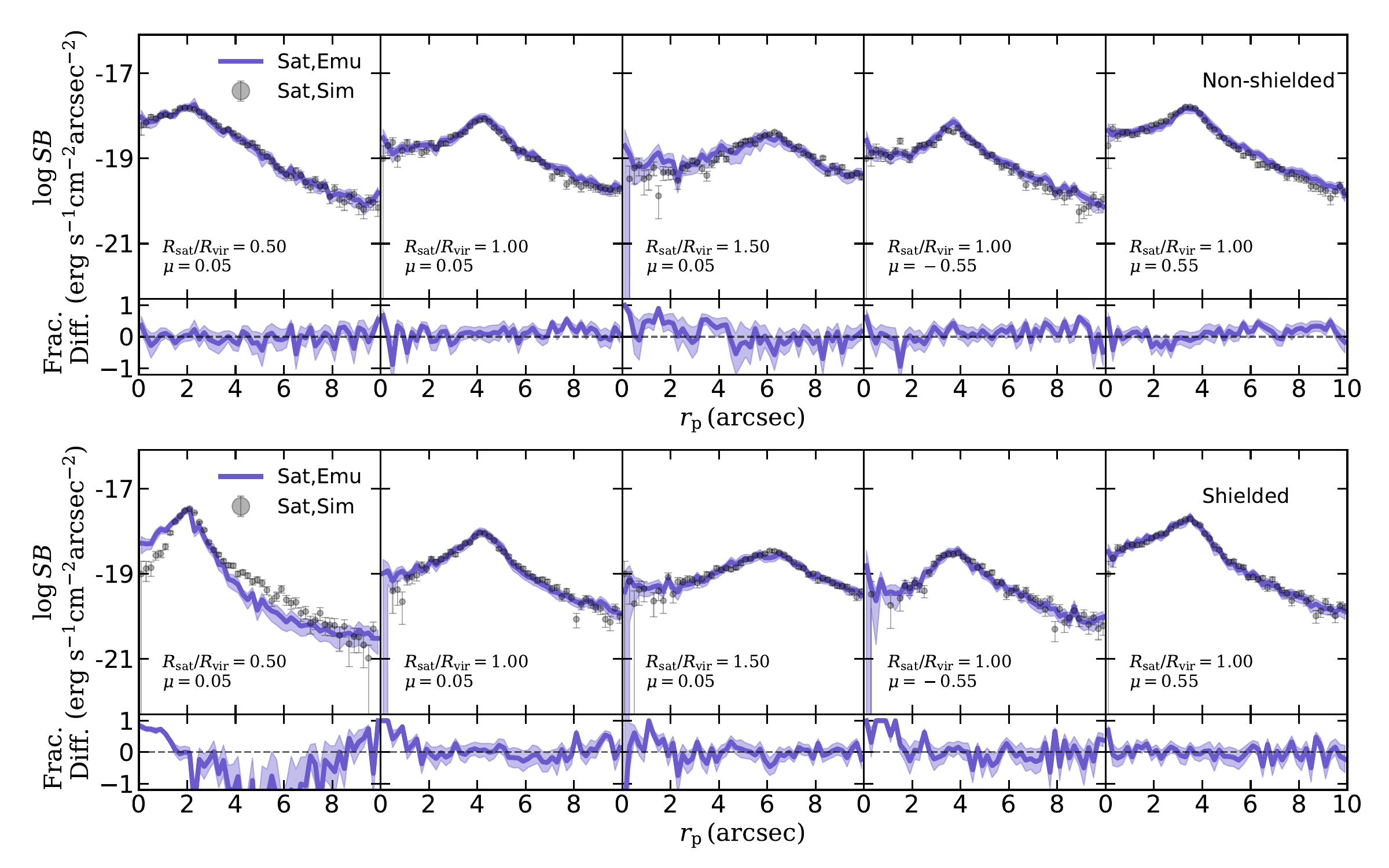}
    \caption{Similar to Fig.~\ref{fig:slhdtest_cc}, but for emulator performance for satellite sources. In the top row, the model parameters relevant for satellite sources are listed, with the rest model parameters fixed at $\log (M_{\rm h}/\Msun)=11$, $z=4$, $\log c_0=0.6$, and $\log f_{\rm esc}=-1.0$. 
    The bottom row is similar except for $\log c_0=1.1$.}
    \label{fig:test_sate}
\end{figure*}

We run RT simulations for the $N=180$ training models in Section \ref{sec:SLHD} and calculate their $\Lya$ SB profiles. 
By interpolating these simulated profiles from training models, we can efficiently obtain $\Lya$ SB profiles for any model without running RT simulations. We adopt the Gaussian Process Regression (GPR) to carry out the interpolation. 

GPR is a non-parametric regression technique, assuming a multivariate Gaussian distribution of the targeting quantity $(y_1, y_2, ..., y_{N})^T \equiv \bm{y}$, 
\begin{equation}
    \mathcal{L} (\bm{y}) = 
    \frac{1}{(2\pi)^{N/2} |C|^{1/2}}
    \exp \left[ -\frac{1}{2} 
    (\bm{y}-\bm{\tilde{y}})^{\rm T} \mathcal{C}^{-1} (\bm{y}-\bm{\tilde{y}})\right].
    \label{eq:MultiGaussian}
\end{equation}
In our case, the quantities in $\bm{y}$ are $\Lya$ SB values of the 180 training profiles at a fixed projected radius. 
Every component in vector $\bm{\tilde{y}}$ takes the median value of $y_1$, $y_2$, ..., and $y_{N}$. 
The covariance matrix $\mathcal{C}$ of this distribution is modelled by the so-called kernel function, such as a Gaussian function or Mat$\rm \Acute{e}$rn series \citep{Rasmussen_2006}. 
In this paper, we use the first-order Mat$\rm \Acute{e}$rn kernel to build emulators, 
\begin{equation}
    \mathcal{C}_{ij} = \sigma^2 \prod_{k=1}^d  \left( 1 + \frac{ \sqrt{3} \left|x_k^{(i)} - x_k^{(j)}\right|}{l_k} \right) \exp \left( -\frac{ \sqrt{3} \left|x_k^{(i)} - x_k^{(j)}\right|}{l_k} \right). 
    \label{eq:MaternKernel}
\end{equation}
The parameter $x_k^{(i)}$ is the scaled model parameter following the same definition as in Section \ref{sec:SLHD}. 
For the two hyper-parameters, the amplitude parameter $\sigma$ describes the global variation of $\bm{y}$, and the scale parameter $l_k$ describes the model correlation length for the $k$-th model parameter. 
We further add a noise kernel to the above kernels for all $\Lya$ sources, 
\begin{equation}
    \mathcal{N}_{ij} = \sigma_{{\rm sn},i}^2 \delta_{ij},
    \label{eq:ShotNoiseKernel}
\end{equation} 
where $\sigma_{{\rm sn},i}$ corresponds to the shot noise of the $\Lya$ SB value $y_i$ from RT simulations. 
The shot noise is estimated as the square root of the simulated photon number for a given projected radius bin. 
Including the noise kernel can suppress the contribution of $\Lya$ SB values associated with small photon number counts during training and predicting processes of the emulator. 

To determine the hyper-parameters in the emulator, we need to maximize the likelihood in equation~(\ref{eq:MultiGaussian}), which is called the training of the emulator. We use the \texttt{emcee} package \citep{Foreman-Mackey_2013} to explore the hyper-parameter space and adopt the hyper-parameters with the largest likelihood. 
Once the training is done, we can predict the $\Lya$ SB value at a fixed projected radius for a given model without running RT simulations \citep{Rasmussen_2006}, 
\begin{equation}
    y^{(\rm p)}= \mathcal{C}^{(\rm pt)} 
    \left(\mathcal{C}^{(\rm tt)}\right)^{-1} \bm{y}^{(\rm t)}, 
    \label{eq:GPR_mean}
\end{equation}
where $\bm{y}^{(\rm t)}$ represents the $\Lya$ SB values from the $N$ training models, $y^{(\rm p)}$ represents the predicted (mean) $\Lya$ SB value for a given model, $\mathcal{C}^{(\rm tt)}$ is the $N \times N$ covariance matrix among the training models calculated through equation~(\ref{eq:MaternKernel})-(\ref{eq:ShotNoiseKernel}), and $\mathcal{C}^{(\rm pt)}$ is the $1\times N$ cross-covariance matrix between the predicting model and the training models calculated through equation~(\ref{eq:MaternKernel}), with $i$ and $j$ denoting the predicting model and the training models, respectively. The cross-covariance matrix has no contribution from the shot noise kernel in equation~(\ref{eq:ShotNoiseKernel}). 
The GPR can also estimate the uncertainty of the predicted value \citep{Rasmussen_2006}, 
\begin{equation}
    \sigma^{(\rm p)} = \sqrt{ \mathcal{C}^{(\rm pp)} -
    \mathcal{C}^{(\rm pt)} \left(\mathcal{C}^{(\rm tt)} \right)^{-1} \left( \mathcal{C}^{(\rm pt)} \right)^{\rm T} },
    \label{eq:GPR_var}
\end{equation}
where $\mathcal{C}^{(\rm pp)}$ is simply $\sigma^2$ from equation~(\ref{eq:MaternKernel}) (with $i=j$ denoting the given model).

\subsection{Implementation of the emulators}\label{sec:implementation}
We build an individual emulator at each fixed projected radius $r_{\rm p}$ for central and CGM sources, and at each fixed projected radius $r_{\rm p}$ and each fixed observing direction $\mu$ for satellite sources. 
To address the large dynamical range of $\Lya$ SB values, we remove the $(1+z)^4$ cosmological dimming effect, take the logarithm of $\Lya$ SB values, and normalize them by subtracting the median and dividing the standard deviation of the $\Lya$ SB values at a fixed projected radius. 
After the emulators are constructed, the cosmological dimming effect is incorporated into the emulated $\Lya$ SB profiles. 

The input model parameters with their ranges are $\log (M_{\rm h}/\Msun)=[10, 12]$, $z=[2, 6]$, $\log c_0=[0.0, 1.2]$, and $\log f_{\rm esc} = [-1.5, -0.5]$ for central and CGM sources, and additionally $R_{\rm sat}/R_{\rm vir}=[0.1, 2]$ and $\mu=[-1, 1]$ for satellite sources. 
We replace the satellite position parameter $R_{\rm sat}/R_{\rm vir}$ with the satellite projected position parameter $R_{\rm sp} \equiv R_{\rm sat}\sqrt{1-\mu^2}$ (in $\rm arcsec$) when constructing the emulators for satellite sources. The new parameter $R_{\rm sp}$ can provide better emulating performance because it denotes the projected radius of the peak $\Lya$ SB values. All model parameters are rescaled to be between zero and one. 

As discussed in Section \ref{sec:RT}, the $\Lya$ SB profiles can be categorized into non-shielded and shielded systems. We find that adding information about the types of the systems can improve the emulating performance. To achieve that, we introduce a derived model parameter to the above free model parameters for all $\Lya$ sources, the peak neutral hydrogen density $\log \nHImax$, defined as the maximum value in the neutral hydrogen density profile after the self-shielding calculation. We find that non-shielded and shielded systems have very different values of $\log \nHImax$ (Appendix \ref{sec:transition}). This parameter will also be rescaled as mentioned above. 

In general, the emulators work well for non-shielded and shielded systems. 
But for systems in between, dubbed transitional systems, the emulator performance can become complicated. 
Therefore, in our following presentations, we focus on examining the emulator performance and applications for non-shielded and shielded systems, and discuss those for transitional systems in Appendix \ref{sec:transition}.

\subsection{Performance of the emulators}\label{sec:performace}
\begin{figure*}
	\includegraphics[width=\textwidth]{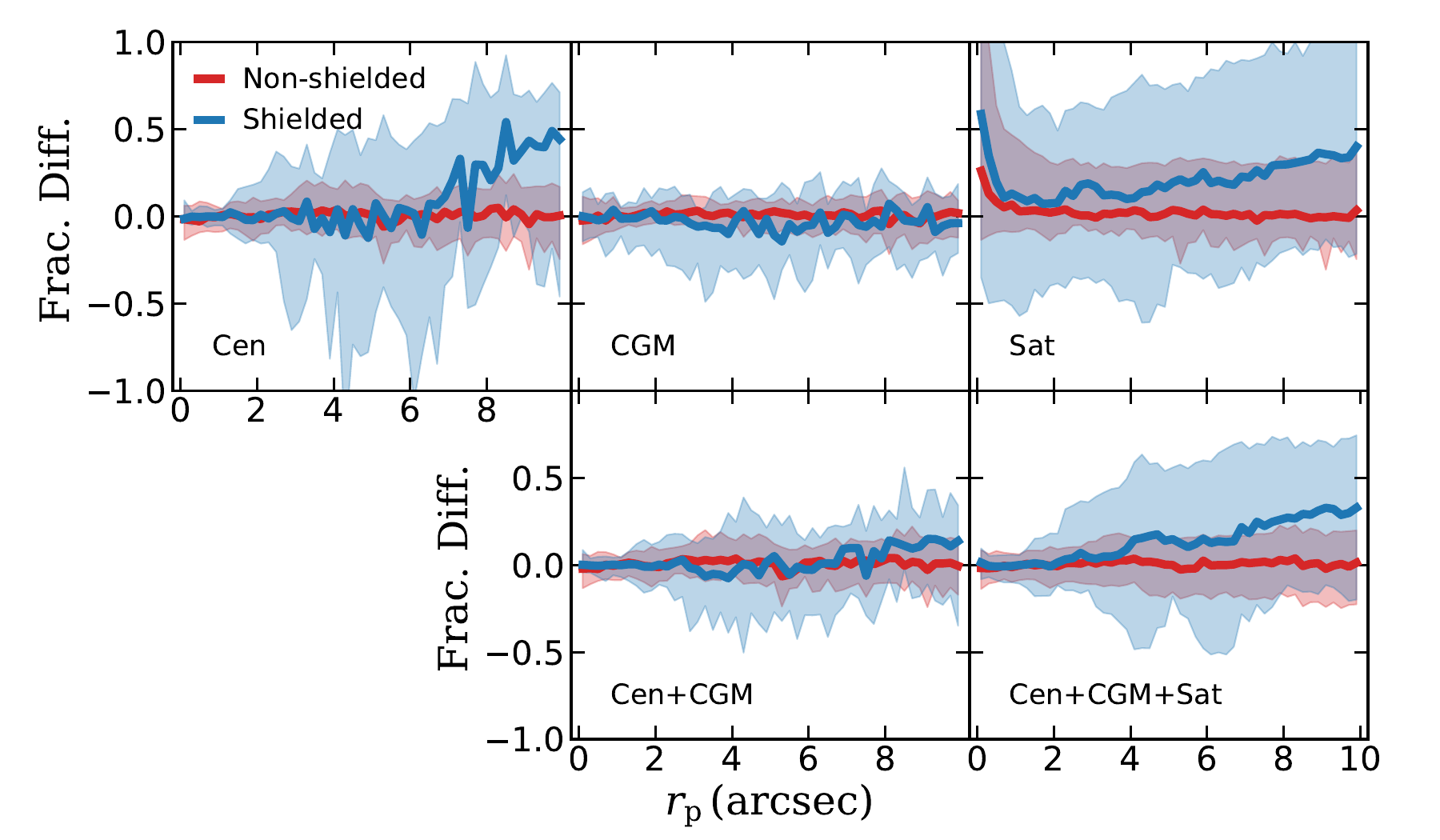}
    \caption{Distribution of the overall accuracy of emulators. Fractional differences between $\Lya$ SB profiles from emulator predictions and from RT simulations, with respect to emulator predictions, are shown for non-shielded systems (red) and shielded systems (blue), and for central, CGM, satellite, Cen+CGM, and Cen+CGM+Sat $\Lya$ sources in the five panels, respectively. For the last case (Cen+CGM+Sat), the satellite-to-central $\Lya$ luminosity ratio is assumed to be 1/2. 
    Solid lines show the median values, while shaded regions mark the range between 16 and 84 percentiles of the distribution, estimated based on the testing models described in Section \ref{sec:performace}. 
    }
    \label{fig:test_global}
\end{figure*}

In this subsection, we examine the performance of the emulators for central, CGM, and satellite sources. 
We further separate the non-shielded and shielded systems given the different shapes of their $\Lya$ SB profiles (Section \ref{sec:RT}). 
Based on our test, we find $\nHImax$ a good indicator of the non-shielded and shielded systems, with $\nHImax<10^{-4} \rm cm^{-3}$ for the former and $\nHImax> 10^{-2} \rm cm^{-3}$ for the latter. 
In between are the transitional systems to be discussed in Appendix \ref{sec:transition}. 
To evaluate the performance, we first create testing models to sample the parameter space. 
Then, we run RT simulations for these testing models to obtain the corresponding $\Lya$ SB profiles and compare them to the predicted ones from the emulators. 

For central and CGM sources, we use the SLHD technique to create testing models by randomly sampling the parameter space. 
Fig.~\ref{fig:slhdtest_cc} shows randomly selected examples of the emulator performance for central and CGM sources in non-shielded systems (top row) and in shielded systems (bottom row). 
In the top panels of each row, the grey points are from RT simulations, and the curves and shaded regions are the means and uncertainties of $\Lya$ SB values predicted by the emulators. 
The emulator means agree well with RT simulations for most cases. 
For cases with slightly larger deviations (Fig.~\ref{fig:slhdtest_cc}, last panel in the top row and last two panels in the bottom row), their $\Lya$ SB profiles tend to be similar to those from transitional systems, with a plateau near the centre followed by a sharp drop at larger radii, which are generally hard to emulate (see Appendix \ref{sec:transition} for more details). 

The bottom panels of each row in Fig.~\ref{fig:slhdtest_cc} show the fractional differences between $\Lya$ SB values from the emulator prediction (mean) and from the RT simulation, with respect to the emulator prediction. 
The shaded regions are the uncertainties of the fractional differences predicted by emulators, which are comparable to the deviation of the fractional differences (solid lines) from 0 for most cases. 
At large radii of shielded systems, the uncertainties are underestimated by emulators for central sources but not for CGM sources. 
In practice, except for systems with $\Lya$ SB profiles resembling those of transitional systems (e.g. the last two panels in the bottom row of Fig.~\ref{fig:slhdtest_cc}), such underestimations for the central sources would have no significant effect, because the CGM sources start to dominate at those radii in typical shielded systems (e.g. the first three panels in the bottom row of Fig.~\ref{fig:slhdtest_cc}). 
In general, the emulator of the CGM sources performs better than that of the central sources. This is expected, since the $\Lya$ SB profiles of the CGM sources vary more slowly with model parameters (except for halo mass) than those of the central sources (see Fig.~\ref{fig:trend_cc}).

Fig.~\ref{fig:test_sate} show the emulator performance for satellite sources with chosen $R_{\rm sat}/R_{\rm vir}$ and $\mu$ as labelled in each panel, in a non-shielded system (top row) and in a shielded system (bottom row). 
While the $\Lya$ luminosity $L_{\rm sat}$ of the satellite is irrelevant for the emulator performance test, for illustration purpose we fix it to be $10^{43} {\rm erg\, s^{-1}}$.
The positions and the shapes of the peak features of $\Lya$ SB profiles are generally well recovered by the emulators. 
The uncertainties predicted by the emulators are comparable to the deviation of the emulator predictions from the simulated values. 

To understand the emulator performance over the whole parameter space for different $\Lya$ sources, we run RT simulations for 92 testing models randomly sampled with the SLHD technique in the parameter space of $\log M_{\rm h}$, $z$, $\log c_0$, $\log f_{\rm esc}$, and $R_{\rm sat}/R_{\rm vir}$. 
Specifically for satellite sources, we further divide $\mu$ into 20 bins uniformly distributed between $-$1 and 1 to construct testing $\Lya$ SB profiles. 
Then we calculate the fractional differences of $\Lya$ SB profiles between emulator predictions and RT simulations for each testing model. 
Fig.~\ref{fig:test_global} shows the distribution of such fractional differences, for different $\Lya$ sources in different panels, and for non-shielded and shielded systems in red and blue. 
The top three panels show the results for central, CGM, and satellite sources, respectively. 
We show results of combined $\Lya$ SB profiles, for the combination of central and CGM sources (Cen+CGM) in the bottom left panel, and the combination of central, CGM, and satellite sources (Cen+CGM+Sat) in the bottom right panel (assuming the satellite-to-central $\Lya$ luminosity ratio $L_{\rm sat}/L_{\rm cen}$ to be 1/2). 
The solid line shows the median of the distribution while the shaded region shows the $1\sigma$ range. 

In general, the non-shielded systems have better emulator accuracy than shielded systems, as a result of more training profiles being non-shielded systems and their $\Lya$ SB profiles being smoother. 
The emulators of central sources tend to overestimate $\Lya$ SB values at large radii in shielded systems. 
But in practice, the overestimation is suppressed after combining central and CGM sources together as shown in the bottom left panel. 
Similarly, the overestimation of satellite emulators in shielded systems can also be suppressed after combining all three sources (as shown in the bottom right panel). 
We notice that the overestimation at large radii is less suppressed in shielded systems after the combination. 
However, given that the uncertainties from observations at those radii are typically large and that the satellite luminosities are typically below our assumed values, we do not further optimize the emulators in this study. 
Using the $1\sigma$ value as a measure for the emulator performance, we calculate the accuracy of emulators for Cen+CGM sources and Cen+CGM+Sat sources, in non-shielded systems and in shielded systems, respectively, as listed in Table \ref{tab:emu_acc}. Combining non-shielded and shielded systems, a simple arithmetic mean of the emulator accuracy is $\sim 17$ per cent for Cen+CGM sources and $\sim 26$ per cent for Cen+CGM+Sat sources. 
\begin{table}
	\centering
	\caption{Emulator accuracy for different $\Lya$ sources in non-shielded and shielded systems.
    }
	\label{tab:emu_acc}
	\begin{tabular}{c c r r} 
		\hline
		    &  & Cen$+$CGM & Cen$+$CGM$+$Sat  \\
		\hline
            \multirow{3}{*}{Non-shielded} & innermost radius & 9.8\% & 11.2\% \\
		                                 & outermost radius & 15.9\% & 21.2\% \\
                                          & averaged over radii & 12.0\% & 15.3\% \\
            \hline
            \multirow{3}{*}{Shielded} & innermost radius & 6.5\% & 8.8\% \\
		                           & outermost radius & 34.6\% & 47.1\% \\
                                      & averaged over radii & 22.4\% & 36.7\% \\
		\hline
    \end{tabular}
    \raggedright
    {\it Note.} The emulator accuracy is measured by the $1\sigma$ value of the fractional difference between $\Lya$ SB profiles from emulator predictions and from RT simulations, obtained with the testing models described in Section \ref{sec:performace}. For each combination of system type and $\Lya$ sources, the emulator accuracy is evaluated at the innermost radius bin (0--0.2 arcsec), at the outermost radius bin (9.8--10.0 arcsec), and over all 50 radius bins (0--10 arcsec), respectively. For the Cen+CGM+Sat cases, the satellite-to-central $\Lya$ luminosity ratio is assumed to be 1/2. 
\end{table}

%% file: sec_application.tex
\section{Application Test of the Emulators} \label{sec:application}

\begin{figure*}
	\includegraphics[width=\textwidth]{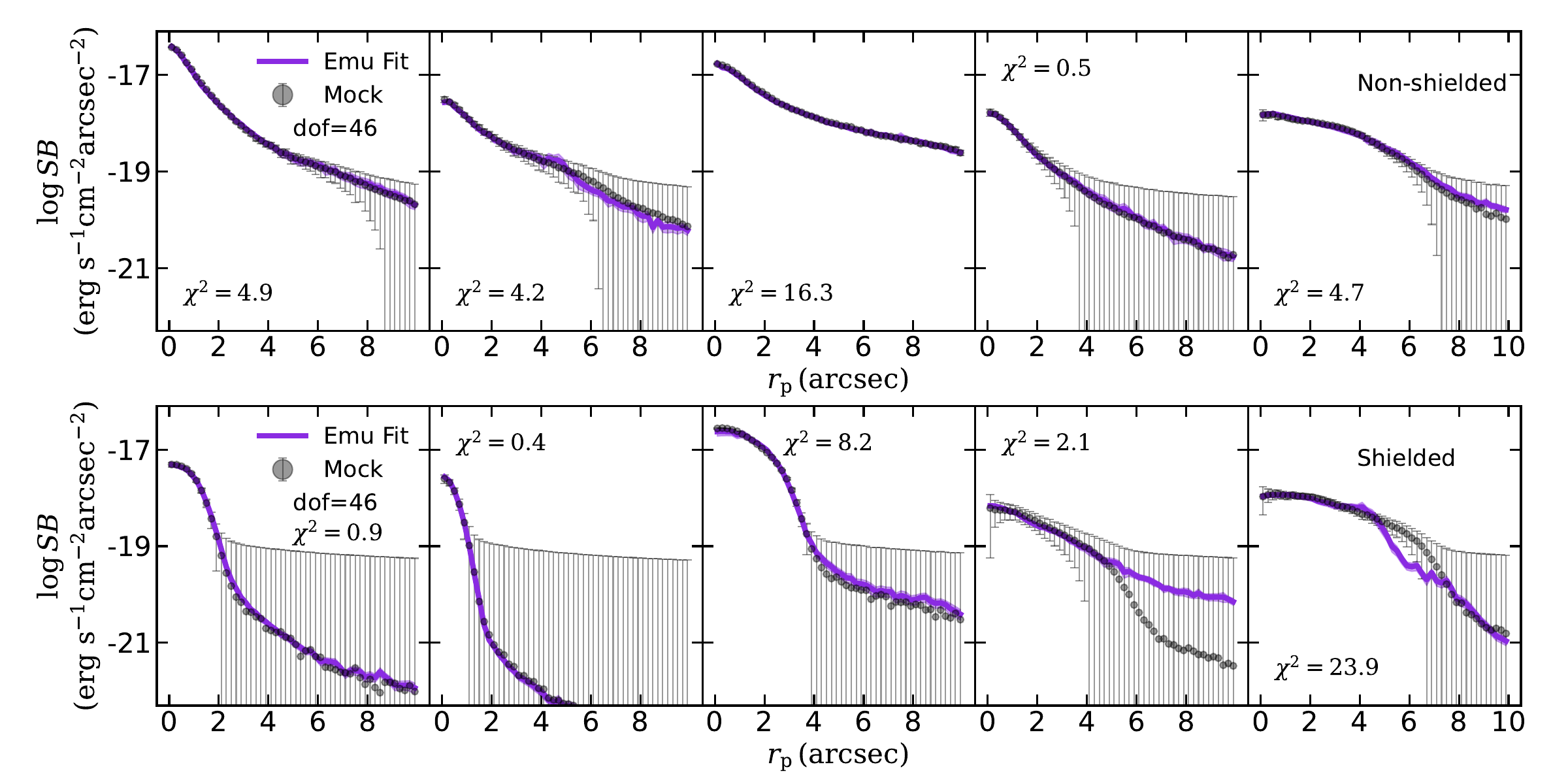}
    \caption{Examples of fitting mock $\Lya$ SB profiles with emulators for Cen+CGM sources in non-shielded systems (top row) and shielded systems (bottom row). 
    The points are the mock $\Lya$ SB profiles constructed from RT simulations, with error bars calculated based on 10-hour MUSE observations. 
    The coloured lines and shaded regions show means and uncertainties predicted by the emulators for the best-fitting models. 
    The free parameters are $\log M_{\rm h}$, $\log c_0$, $\log f_{\rm esc}$, and $\log \nHImax$, and the redshift $z$ is fixed at the underlying truth. In each panel, the best-fitting $\chi^2$ value (incorporating uncertainties from mock observations and those predicted by emulators) is listed, with degrees of freedom (dof) being 46. 
    }
    \label{fig:fit_slhdtest}
\end{figure*}

\begin{figure*}
	\includegraphics[width=\textwidth]{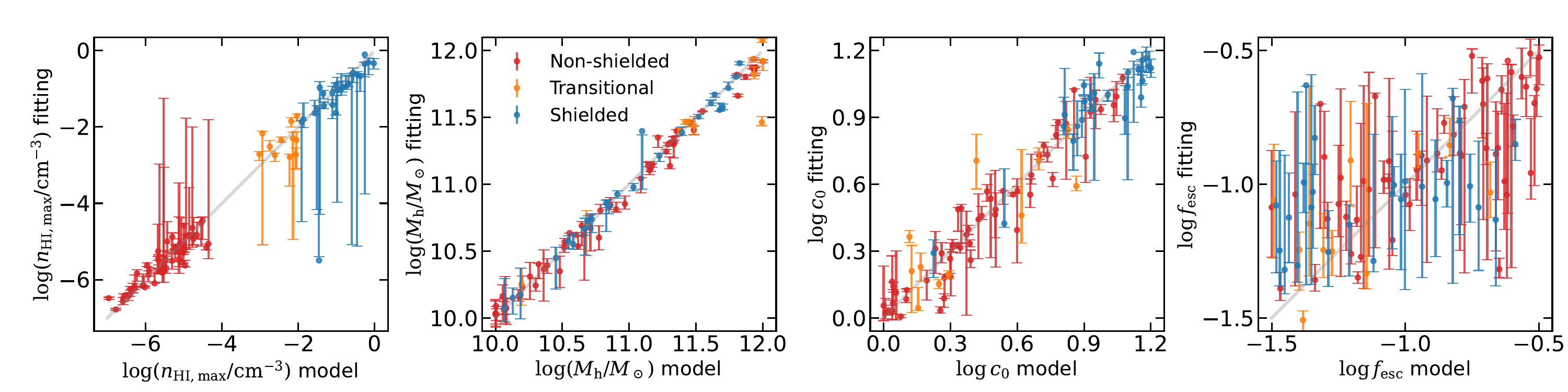}
    \caption{Results of parameter recovery from fitting $\Lya$ SB profiles of 92 randomly selected models for Cen+CGM sources.
    Best-fitting model parameters (best-fitting values with the marginalized 1$\sigma$ ranges; $y$-axes) are compared with those used in generating the mock $\Lya$ SB profiles ($x$-axes), with the diagonal line in each panel marking the equality relation.
    The red, orange, and blue colours correspond to non-shielded, transitional, and shielded systems, respectively. 
    }
    \label{fig:param_recovery_cc}
\end{figure*}

In this section, aiming at applying the emulators to observational data, we use them to fit mock $\Lya$ SB profiles and constrain model parameters. 
We discuss the fitting procedure and investigate the recovery of model parameters. 

We construct mock $\Lya$ SB profiles based on our model. First, we run RT simulations on 92 testing models selected by the SLHD technique, and obtain their $\Lya$ SB profiles convolved with a typical MUSE PSF of $\rm FWHM = 0.7\, arcsec$. 

Next, we add noises to the simulated profiles mimicking MUSE observational conditions. 
For each system, we first find the observed wavelength range that covers 90 per cent of the simulated $\Lya$ photons \citep{Wisotzki_2016}, excluding 5 per cent of photons on either tail of the distribution. 
Then, we calculate the sky surface brightness by integrating the sky background spectrum, provided by \href{https://www.eso.org/observing/etc/bin/gen/form?INS.NAME=MUSE+INS.MODE=swspectr}
{MUSE exposure time calculator}, within the selected wavelength range. 
We set the exposure time to be $36000\, \rm s$ and the overall telescope and instrument efficiency to be $0.9$. 
Then, we obtain the number of photons received from both the sky and the source, and assign the Poisson noise to the $\Lya$ SB value at each radius. 

To fit the model to a mock $\Lya$ SB profile, we construct the likelihood function 
\begin{equation}
    \mathcal{L}_{\rm fit} = \prod_{i=1}^{ N_{r_{\rm p}} } 
    \frac{1}{\sqrt{2\pi} \sigma_i} 
    \exp \left[ -\frac{ \left( y_i^{(\rm o)} - y_i^{(\rm p)} \right)^2}{2 \sigma_i^2} \right], 
    \label{eq:GaussianLikelihood}
\end{equation}
where $N_{r_{\rm p}}$ is the number of projected radius bins, $y^{(\rm o)}_i$ and $y^{(\rm p)}_i$ are the $\Lya$ SB values of the $i$th projected radius bin from mock observations and emulators, respectively, and $\sigma_i$ is the corresponding uncertainty. For uncertainties, we include those from the mock observations ($\sigma^{(\rm o)}_i$) and from the emulators [$\sigma^{(\rm p)}_i$; equation~(\ref{eq:GPR_var})], added in quadrature. Notice that all the terms in the likelihood function are calculated in the linear space of $\Lya$ SB values instead of the logarithmic space. 
We use the \texttt{emcee} package \citep{Foreman-Mackey_2013} to explore the model parameter space. 

Below we show our modelling results for two cases, Cen+CGM and Cen+CGM+Sat.  
Since the redshifts of LAHs are usually known from observations, we fix the redshifts in modelling $\Lya$ SB profiles. 
We treat the peak neutral hydrogen density $\nHImax$ as a free parameter as well, given our emulator setup (Section \ref{sec:implementation}). 

We first focus on modelling $\Lya$ SB profiles with central and CGM sources. In Fig.~\ref{fig:fit_slhdtest}, we show fitting examples for non-shielded systems in the top row and for shielded systems in the bottom row. 
The grey points are the mock $\Lya$ SB profiles combining central and CGM sources based on randomly selected models in Fig.~\ref{fig:slhdtest_cc}, and the coloured lines are emulated profiles from the best-fitting models. 
The $\chi^2$ value listed in each panel is calculated by incorporating the uncertainties from mock observations and those predicted by emulators. 
The emulators can well reproduce the mock profiles for both non-shielded and shielded systems. 

To test the recovery performance of model parameters, we fit mock $\Lya$ SB profiles of the 92 randomly selected testing models and compare their best-fitting parameters to the truth values of the testing models. 
The results are shown in Fig.~\ref{fig:param_recovery_cc}, with non-shielded systems in red and shielded systems in blue. 
As a comparison, we also show the results of transitional systems in orange. 

As shown in the first panel, the well constrained $\nHImax$ values are consistent with the underlying truths of the testing models. 
The halo mass $M_{\rm h}$ and the gas concentration $c_0$ are well constrained as well, while the escaping fraction $f_{\rm esc}$ is loosely constrained due to its weak effects on $\Lya$ SB profiles compared to other parameters (Fig.~\ref{fig:trend_cc}). 
The model parameter constraints of transitional systems are sometimes biased as a result of the inadequate emulator accuracy for such systems (see Appendix \ref{sec:transition}). 

\begin{figure*}
	\includegraphics[width=\textwidth]{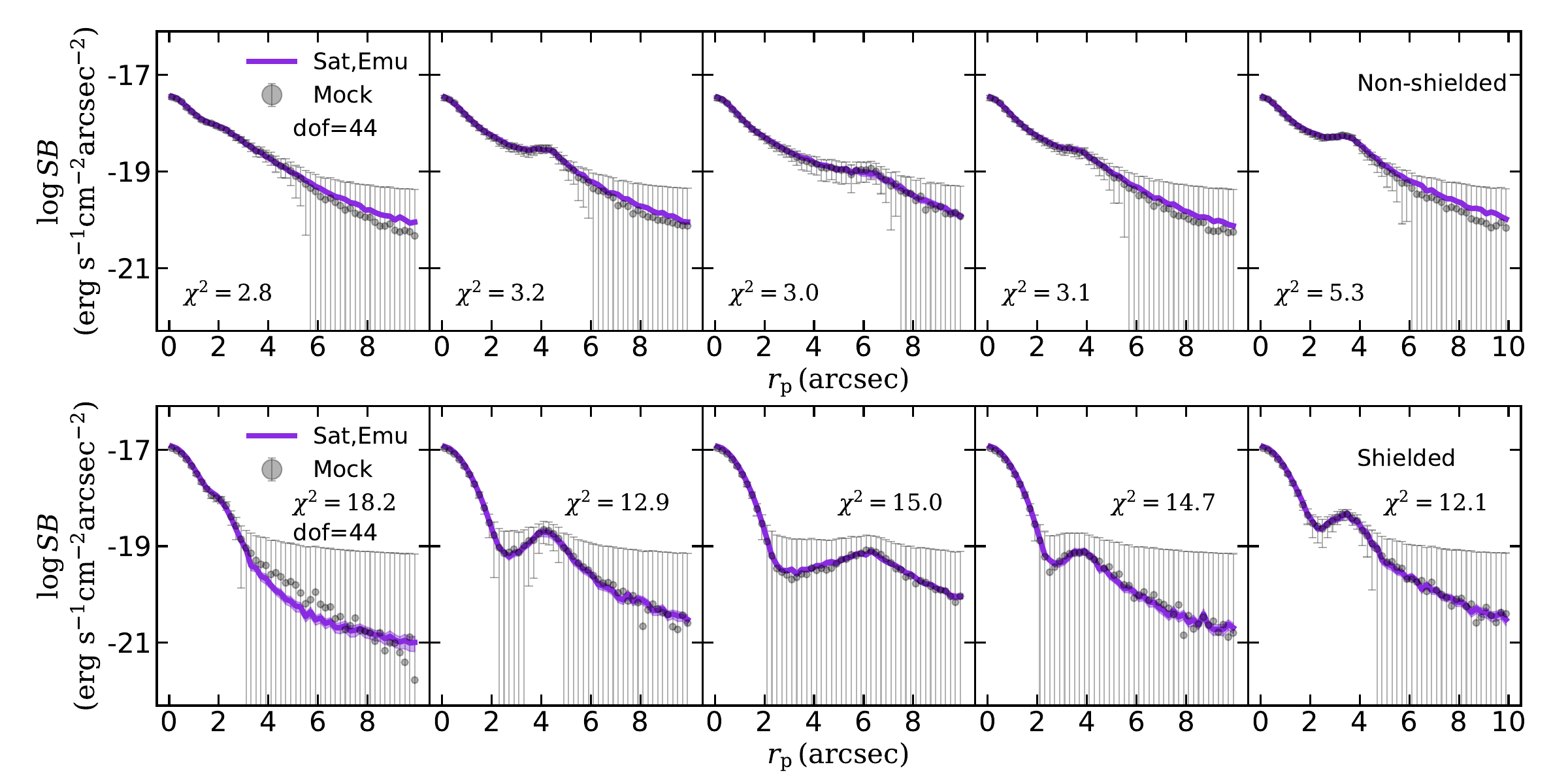}
    \caption{Similar to Fig.~\ref{fig:fit_slhdtest}, but for fitting mock $\Lya$ SB profiles with emulators for Cen+CGM+Sat sources. 
    The free parameters are $\log M_{\rm h}$, $\log c_0$, $\log f_{\rm esc}$, $\log \nHImax$, $\mu$, and $\log f_{\rm sc}$. 
    The redshift $z$ and satellite projected position $R_{\rm sp}$ are fixed at the underlying truths. 
    The resulting degrees of freedom (dof) is 44.}
    \label{fig:fit_gridtest_line_sate}
\end{figure*}

\begin{figure*}
	\includegraphics[width=\textwidth]{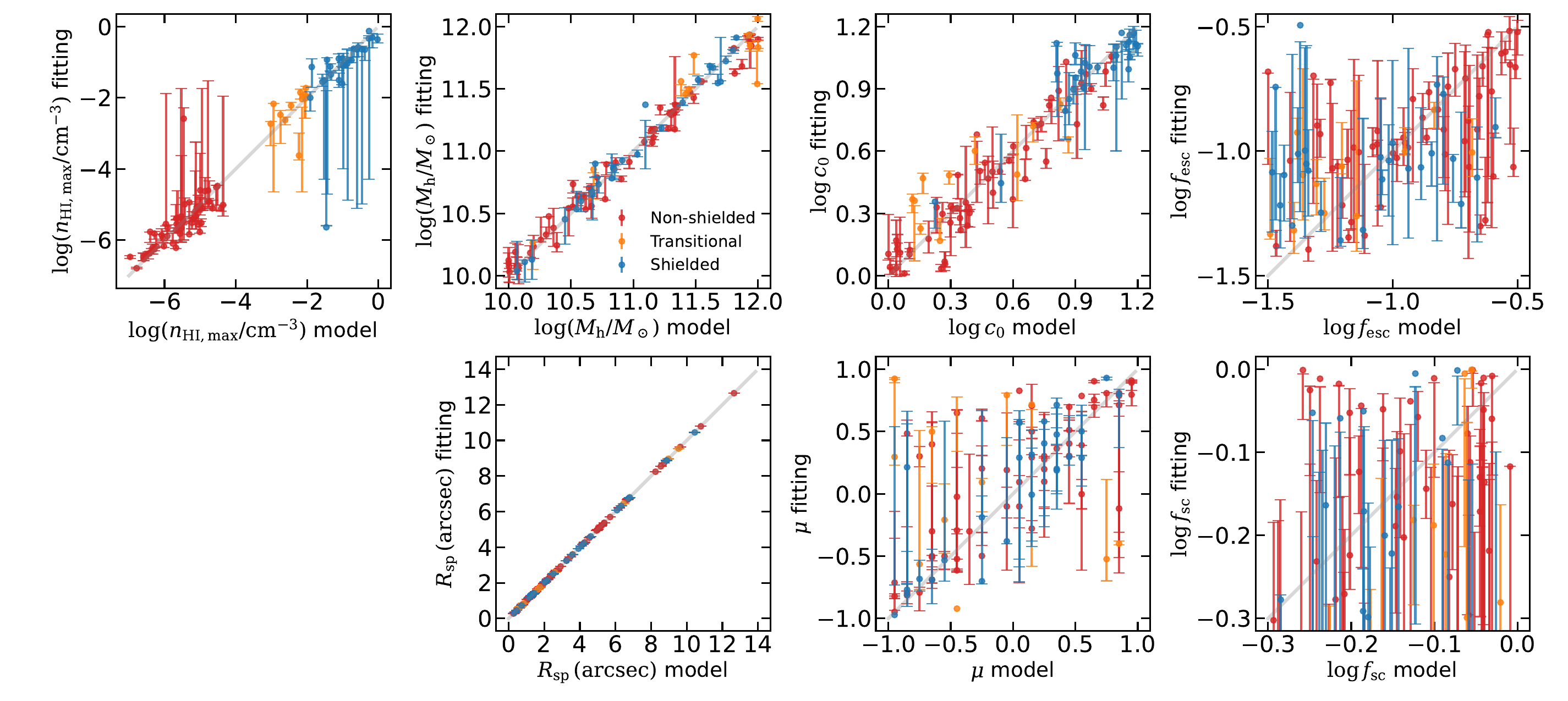}
    \caption{Similar to Fig.~\ref{fig:param_recovery_cc}, but for parameter recovery from fitting $\Lya$ SB profiles of 92 randomly selected models for Cen$+$CGM$+$Sat sources. 
    The projected position of satellite sources, $R_{\rm sp}$, shown in the bottom-left panel, is fixed during the fitting.}
    \label{fig:param_recovery_sate}
\end{figure*}
Now, we discuss modelling results for $\Lya$ SB profiles with the presence of central, CGM, and satellite sources. 
Fig.~\ref{fig:fit_gridtest_line_sate} presents the fitting examples for non-shielded systems in the top row and for shielded systems in the bottom row. 
We set the intrinsic luminosity $L_{\rm sat}$ of a satellite source to be half of that of the central source in each system, which creates a bump feature on the mock profiles. 
To fit $\Lya$ SB profiles with such features, we need three additional model parameters, the projected satellite position $R_{\rm sp}$, the observing direction $\mu$, and the ratio of intrinsic $\Lya$ luminosities between satellite and central sources, $f_{\rm sc}=L_{\rm sat}/L_{\rm cen}$. 
If the satellite source makes a noticeable contribution to the (circularly averaged) $\Lya$ SB profiles, it would show up clearly in the narrow-band image, which enables us to have an accurate determination of $R_{\rm sp}$. 
Therefore, we fix the parameter $R_{\rm sp}$ to the underlying truths in the fitting. For $\mu$, the emulators are built on discrete $\mu$ bins. 
To emulate $\Lya$ SB profiles for an arbitrary $\mu$ value, a linear interpolation is performed between the two emulated values in the neighbouring $\mu$ bins. 
The emulator predictions of the best-fitting models agree well with the mock $\Lya$ SB profiles (Fig.~\ref{fig:fit_gridtest_line_sate}), with both the overall shapes and the bump features adequately recovered. 

Fig.~\ref{fig:param_recovery_sate} shows the parameter recovery from fitting Cen+CGM+Sat sources with the 92 randomly selected testing models. 
We also randomly sample the observing direction $\mu$ between $-1$ and 1 and the satellite-to-central $\Lya$ luminosity ratio parameter $\log f_{\rm sc}$ between $-0.3$ and 0.0 (so that satellites could produce noticeable features on the $\Lya$ SB profiles). 
We fix $R_{\rm sp}$ when carrying out the fitting. 
Similar to the Cen+CGM cases in Fig.~\ref{fig:param_recovery_cc}, the model parameters $\log \nHImax$, $\log M_{\rm h}$, and $\log c_0$ can be well recovered, while the constraints on $\log f_{\rm esc}$ are relatively loose. 
For many cases, the free parameters for satellite sources, $\mu$ and $\log f_{\rm sc}$, are loosely constrained, mainly because satellites do not lead to significant features in the $\Lya$ SB profiles after being circularly averaged. 
In addition, there is degeneracy between $\mu$ and $f_{\rm sc}$. 
At a given $R_{\rm sp}$, with respect to the central source, a satellite on the near side towards the observer with a lower $\Lya$ luminosity has a $\Lya$ SB profile similar to that of a satellite on the far side with a higher $\Lya$ luminosity, which makes it hard to constrain $\mu$ and $f_{\rm sc}$ individually. 
To improve the constraints on satellite-relevant parameters, it would be better to model the two-dimensional $\Lya$ image instead of the one-dimensional, circularly averaged $\Lya$ SB profile. 

In summary, by applying the emulators to fit mock $\Lya$ SB profiles, we find that model parameters can be well recovered for $\log M_{\rm h}$ and $\log c_0$. 
The constraints on $\log c_0$ are largely determined by the shape of the $\Lya$ SB profiles (see Fig.~\ref{fig:trend_cc}).
The tight constraints on $\log M_{\rm h}$ mainly come from matching the amplitude of the $\Lya$ SB profiles (see Fig.~\ref{fig:trend_cc}), which is tied to the $\Lya$ luminosity $L_{\rm cen}$ from the central SFR [equation~(\ref{eq:LyaCentral})]. Scatters in the $L_{\rm cen}$-$M_{\rm h}$ relation are expected to exist and need to be incorporated into a more realistic model, and as a consequence the constraint on $\log M_{\rm h}$ would be relaxed.
For other parameters that are loosely constrained, it is usually caused by their weak influence on $\Lya$ SB profiles or parameter degeneracy. 

%% file: sec_summary.tex
\section{Summary and Discussions} \label{sec:summary} 
In this paper, based on $\Lya$ RT simulations with an analytically parametrized model of gas distribution, we develop emulators for the SB profiles of the extended $\Lya$ emission (on scales of kpc to tens of kpc) around star-forming galaxies. We summarize our main results below. 
\begin{itemize}
    \item[1.] We construct a parametrized model for $\HI$ gas in and around haloes of star-forming galaxies, described by the density, velocity, and temperature profiles. 
    The self-shielding calculation is performed to obtain the neutral hydrogen densities, accounting for ionizing photons from both the central star formation and the UV background. 
    It reveals three distinct types of neutral gas distribution, the non-shielded system with the entire system being highly ionized, the shielded systems with the gas forming a shell of a high neutral fraction that shields the inner part from the UV background, and the transitional system in between. 
    The type of a system depends on model parameters such as halo mass, gas concentration, escaping fraction of ionizing photons from the central galaxy, and redshift. 
    
    \item[2.] Three $\Lya$ sources are considered in our model, including star formation in the central galaxy, recombination in the CGM, and star formation in the satellite galaxy. 
    For a given gas model, we run $\Lya$ RT simulation for each source to obtain the corresponding $\Lya$ SB profile. 
    In general, the $\Lya$ SB profiles of central and CGM sources tend to have similar shapes, with those of central sources being more concentrated and dominated in the inner region. 
    For satellite sources, a bump can show up in the $\Lya$ profile around the projected satellite position. 
    The three types of the neutral gas distribution lead to different features of $\Lya$ SB profiles, in particular for central and CGM sources. 
    The $\Lya$ SB profiles of non-shielded systems approximately drop exponentially. 
    For shielded systems, the $\Lya$ SB profiles drop fast at the centre, creating a concentrated core followed by a roughly exponential tail. 
    For transitional systems, the central core becomes much more extended compared to shielded systems. 
    
    \item[3.] Emulators are constructed for $\Lya$ SB profiles for each of the $\Lya$ sources based on GPR, which can efficiently predict $\Lya$ SB profiles for given model parameters. 
    The SLHD technique is adopted to uniformly sample the model parameter space to obtain the models for training the emulators. 
    With only 180 training models, we demonstrate that the $\Lya$ SB profiles predicted by the emulators can properly reproduce those from RT simulations for most cases. 
    The emulators function well for non-shielded and shielded systems, with the overall accuracy at the level of $\sim 14$ per cent and $\sim 30$ per cent  (Table \ref{tab:emu_acc}), respectively. 
    For transitional systems, they occupy a small fraction of the parameter space. 
    Their $\Lya$ SB profiles are sensitive to the change of model parameters, which makes it challenging to build emulators with good performance. 
    
    \item[4.] We test the emulators by applying them to fit mock $\Lya$ SB profiles. 
    The mock $\Lya$ SB profiles are produced from a set of testing models, which are different from the training models. 
    The emulators can fit the mock $\Lya$ SB profiles well, with constraints on parameters like halo mass and gas concentration agreeing well with the underlying truths. 
\end{itemize}

While the model we present in this paper is adequate for the pilot study of developing the emulator framework, more physical ingredients can be added to improve the model. 
First, one important ingredient is the gas outflow, which is ubiquitous in star-forming galaxies and affects the $\Lya$ SB profiles \citep[e.g.][]{Zheng+2002a, Dijkstra_2008b, Dijkstra_2012}. 
The turbulent motion of gas is simply treated in our model by introducing a scale-independent effective temperature, while the turbulence can couple the gas density and velocity on different scales \citep[e.g.][]{Kimm_2019, Kimm_2022, Chen_2023, Munirov_2023}. 
The dust in the CGM is another ingredient to be considered \citep[e.g.][]{Hayes_2013, Duval_2014, Laursen_2013}. 
The collisional excitation/ionization can also be included in the model, which can affect the neutral hydrogen density profiles (through self-shielding calculation) and the intrinsic $\Lya$ emissivity (i.e. cooling radiation). 
All these ingredients collectively shape the $\Lya$ spectra and modify the $\Lya$ SB profiles. 
Second, the density, velocity, and temperature distributions of the CGM gas are far from being uniform and isotropic, as a result of processes and structures related to the gas accretion and outflow, leading to anisotropic $\Lya$ emission \citep[e.g.][]{Laursen_2007, Verhamme_2012, Zheng_2014, Smith_2022, Blaizot_2023}. 
Finally, the IGM attenuation suppresses the $\Lya$ emission and reshapes the $\Lya$ emission properties \citep[e.g.][]{Byrohl_2020}. 
Incorporating these ingredients in the model can help us better understand $\Lya$ SB profiles and $\Lya$ emission in general. 

While the performance of our emulators is reasonable, there are possible ways to further improve it. In general, it helps the emulator performance by increasing the number of training models, combining different kernel functions, and re-parametrizing the model. 
In our work, we find that the sharp change in $\Lya$ SB profiles between non-shielded and shielded systems makes it difficult to obtain uniformly accurate emulators across the whole parameter space. 
With good descriptions of boundaries that separate non-shielded, shielded, and transitional systems in the parameter space, it is possible to build separate emulators for each type of the systems to improve the performance. 
Applying dimensional reduction techniques (e.g. Principle Components Analysis) before training emulators allows us to extract important features of $\Lya$ SB profiles, potentially making the emulators more robust and reducing the number of emulators to be trained. 
Our current emulators are separately built at each fixed radius, but in principle, the correlation of $\Lya$ SB values among different radii can also be accounted for (e.g. with the so-called multi-scale emulation technique; \citealt{Dumerchat_2024}) to further improve the emulator accuracy. 
While we focus on $\Lya$ SB profiles in this work, our study can be extended to emulate the IFU data cube of the $\Lya$ emission. 

Our emulators can be applied to fit $\Lya$ SB profiles based on galaxy formation simulations  \citep[e.g.][]{Lake_2015, Byrohl_2021, Byrohl_2023}. 
By comparing the gas distribution from our modelling results and that from galaxy formation simulations, we can validate our modelling choices, understand the connections between model parameters and CGM properties, and provide valuable insights for future model developments. 
We will apply our emulators to model $\Lya$ SB profiles of individual star-forming galaxies from IFU observations \citep[e.g.][]{Leclercq_2017} and stacked $\Lya$ SB profiles from narrow-band imaging observations \citep[e.g.][]{Kikuta_2023}, which will not only probe the CGM gas but also advance our understanding of the extended $\Lya$ emission.

%% file: sec_appendix.tex
\appendix

\section{Transitional Systems} \label{sec:transition}
\begin{figure}
	\includegraphics[width=\columnwidth]{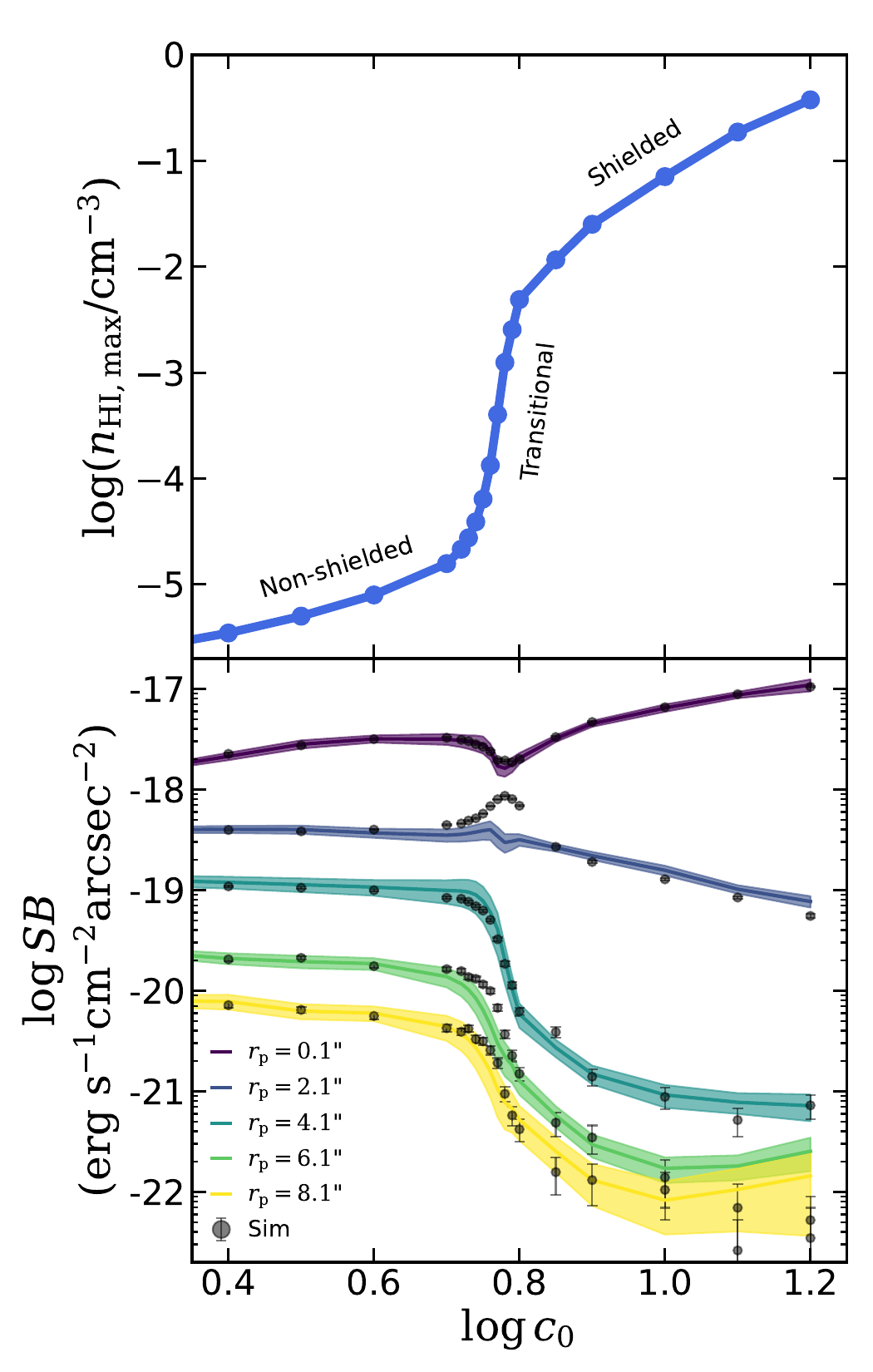}
    \caption{An example of the responses of $\nHImax$ and $\Lya$ SB values to gas concentration $\log c_0$, illustrating the transition from non-shielded, to transitional, and to shielded systems. 
    In the bottom panel, $\Lya$ SB values at five radii are shown. 
    The points are from RT simulations, while the curves and shaded regions are the means and uncertainties predicted by the emulators.
    }
    \label{fig:cg_sbp_pt}
\end{figure}
As discussed in the main text, the systems in our model can be categorised into non-shielded, shielded, and transitional systems. The transitional systems have special features on their $\Lya$ SB profiles and bring challenges to build accurate emulators. In the appendix, we provide a systematic discussion on the transitional systems. 

The transitional systems in our paper are defined to have $\log (\nHImax/\rm cm^{-3})$ between $-4$ and $-2$, where $\nHImax$ is the maximum neutral hydrogen density of the system after self-shielding calculations. 
Among the 180 training models, only 23 of them are transitional systems. 
For transitional systems, a small change in model parameters leads to a big change in neutral hydrogen density profiles. 
The corresponding $\Lya$ SB profiles also sensitively depend on model parameters, typically showing a plateau near the centre followed by a sharp drop at large radii. 
The top panel of Fig.~\ref{fig:cg_sbp_pt} shows the dependence of $\log \nHImax$ on $\log c_0$ for testing models, with other model parameters fixed at $\log (M_{\rm h}/\Msun)=11$, $z=4$, and $\log f_{\rm esc}=-1$. 
As the gas concentration increases, the system in this set of testing models transforms from non-shielded, to transitional, and to shielded. 
At $\log c_0 \sim 0.7$, the value of $\log \nHImax$ increases sharply with $\log c_0$, signalling the system to be transitional, and a gas shell with high neutral hydrogen density forms (see Fig.~\ref{fig:trend_cc}). 
The bottom panel of Fig.~\ref{fig:cg_sbp_pt} shows the dependence of $\Lya$ SB values (of a central source) on $\log c_0$ at different fixed radii. 
The black points are from RT simulations of the testing models and the coloured lines are the emulator predictions. 
The $\Lya$ SB values display complicated behaviours around $\log c_0 \sim 0.7$, oscillating with $\log c_0$ at small radii and sharply decreasing with $\log c_0$ at large radii. 
Apparently, the emulators for transitional systems are not as accurate as those for non-shielded and shielded systems. 

\begin{figure*}
	\includegraphics[width=\textwidth]{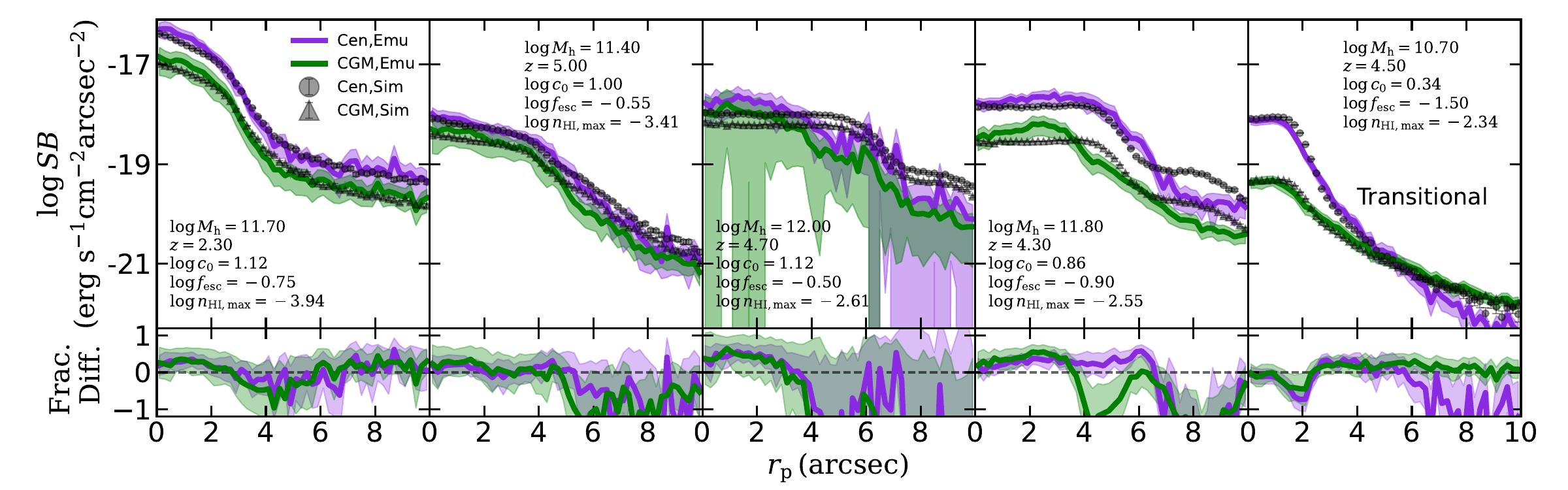}
    \caption{Similar to Fig.~\ref{fig:slhdtest_cc}, but for emulator performance for central and CGM sources in transitional systems.}
    \label{fig:ptslhdtest}
\end{figure*}

\begin{figure*}
	\includegraphics[width=\textwidth]{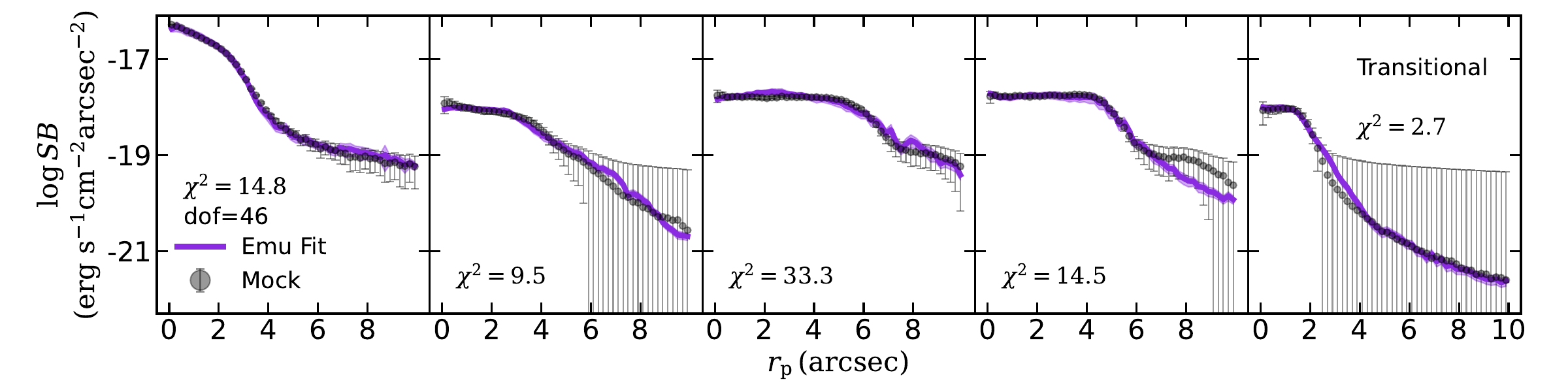}
    \caption{Similar to Fig.~\ref{fig:fit_slhdtest}, but for fitting mock $\Lya$ SB profiles with emulators for Cen+CGM sources in transitional systems.}
    \label{fig:fit_ptslhdtest}
\end{figure*}

\begin{figure*}
	\includegraphics[width=\textwidth]{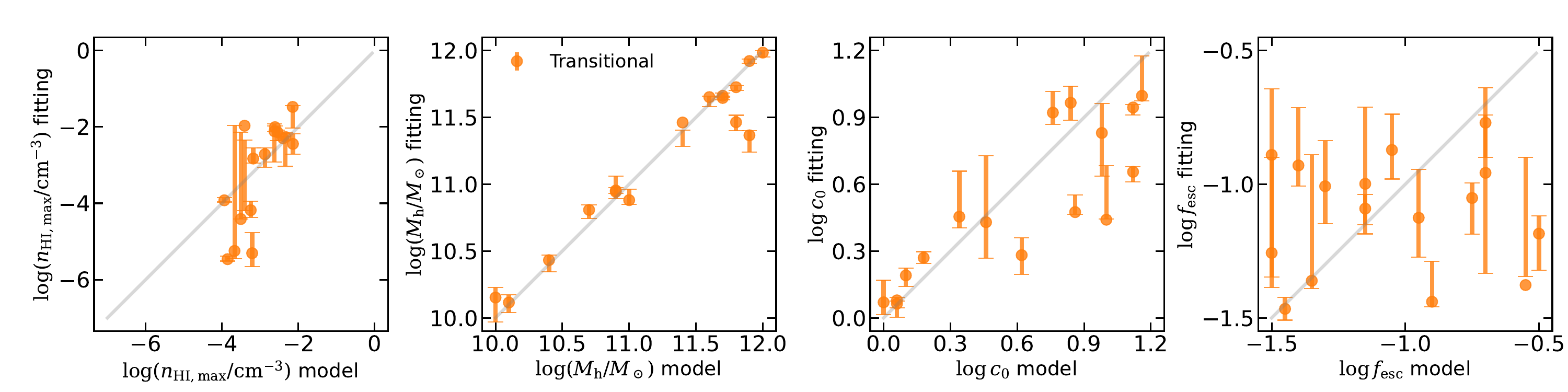}
    \caption{Similar to Fig.~\ref{fig:param_recovery_cc}, but for parameter recovery from fitting $\Lya$ SB profiles for Cen+CGM sources in transitional systems.}
    \label{fig:param_recovery_cc_pt}
\end{figure*}

Fig.~\ref{fig:ptslhdtest} shows the emulator performance for central and CGM sources in selected transitional systems (similar to Fig.~\ref{fig:slhdtest_cc}). 
The $\Lya$ SB profiles from emulators have large deviations compared to those from RT simulations. 
By selecting 16 models of transitional systems with $\log (\nHImax/\rm cm^{-3})$ uniformly distributed between $-4$ and $-2$, we estimate the emulator accuracy for Cen+CGM sources to be $\sim 41$ per cent (i.e. the $1\sigma$ value of the fractional difference).  
The uncertainties predicted by emulators are larger compared to those of non-shielded and shielded systems, indicating complex dependence of $\Lya$ SB profiles on model parameters. 
For satellite sources in transitional systems, if they locate inside the self-shielding radii, the emulator performance for corresponding $\Lya$ SB profiles resembles that of the central sources in such systems. 
If they locate outside, their emulator performance is not much different from that of satellite sources in non-shielded or shielded systems. 
Therefore, we do not specifically present the emulator performance for satellite sources here. 
To improve the emulator performance for the transitional systems, it requires more training models to sample the relevant parameter space. 
In the parameter space explored in this work, only a small fraction ($\sim 13$ per cent) of the models corresponds to transitional systems. 
If future comparisons to observed $\Lya$ SB profiles hint the existence of transitional systems, more effort would be needed to improve the accuracy of emulators for such systems. 

To test the modelling and parameter recovery for transitional systems, we construct mock $\Lya$ SB profiles using the above 16 testing models for Cen+CGM sources as discussed in Section \ref{sec:application}. 
Fig.~\ref{fig:fit_ptslhdtest} shows the best-fitting model profiles together with the mock profiles for five examples (similar to Fig.~\ref{fig:fit_slhdtest}), while Fig.~\ref{fig:param_recovery_cc_pt} shows the parameter recovery for all 16 models (similar to Fig.~\ref{fig:param_recovery_cc}). 
Although the best-fitting $\Lya$ SB profiles agree well with the mock profiles in Fig.~\ref{fig:fit_ptslhdtest}, in a few cases, the modelling leads to biased constraints on certain model parameters (Fig.~\ref{fig:param_recovery_cc_pt}), as a result of the limited accuracy of the emulators. 
In other cases, certain model parameters can still be well recovered (e.g. $\log M_{\rm h}$), which is expected given that $\Lya$ SB profiles for transitional systems are sensitive to those model parameters. 
In general, the emulators for transitional systems can still be applied to provide useful constraints on model parameters, but it requires special attention to compare the emulator predictions with RT simulation results. 